\documentclass[twocolumn,superscriptaddress,amsmath,amssymb,aps,prx,nofootinbib,longbibliography]{revtex4-1}

\usepackage{amsmath}
\usepackage{amssymb}
\usepackage{amsthm}
\usepackage{mathtools}
\usepackage{mathdots}
\usepackage{amsfonts}
\usepackage{bbm}
\usepackage{xr}
\usepackage{bbold}
\usepackage{newpxtext}
\usepackage{epsfig,color,dsfont,upgreek,physics}
\usepackage{mathrsfs}
\usepackage{multirow}
\usepackage{graphicx}
\usepackage{dcolumn} 
\usepackage{bm} 
\usepackage{float}  
\usepackage{longtable}
\usepackage{ulem}  
\usepackage{hyperref}
\allowdisplaybreaks

\renewcommand\[{\begin{equation}}
\renewcommand\]{\end{equation}} 

\footnotetext{J. H. and D. S. contributed equally to this work.}

\begin{document}

\title{Supercurrent Diode Effect in Helical Superconductors}

\author{Jaglul Hasan$^*$}
\affiliation
{Department of Physics, University of Wisconsin-Madison, Madison, Wisconsin 53706, USA}

\author{Daniel Shaffer$^*$}
\affiliation
{Department of Physics, University of Wisconsin-Madison, Madison, Wisconsin 53706, USA}

\author{Maxim Khodas}
\affiliation{The Racah Institute of Physics, The Hebrew University of Jerusalem, Jerusalem 91904, Israel}

\author{Alex Levchenko}
\affiliation
{Department of Physics, University of Wisconsin-Madison, Madison, Wisconsin 53706, USA}

\begin{abstract}
In this work, we explore the generalities of the supercurrent diode effect. As an illustrative example, we examine a model of a two-dimensional superconductor with Rashba-type spin-orbit coupling under an in-plane magnetic field and in the clean limit, which realizes a helical phase. First, we utilize Ginzburg-Landau phenomenology to derive a general formula for the diode efficiency. This is achieved by incorporating higher gradient terms in the Lifshitz invariants, which are responsible for the nonreciprocal superflow. Subsequently, we validate these results through microscopic diagrammatic computation and further estimate correction terms arising from interband pairing correlations. We provide a detailed comparison to prior investigations of this problem conducted within the framework of the quasiclassical approximation based on the Eilenberger equation.
\end{abstract}
\date{April 25, 2024}
\maketitle


\section{Introduction}

The supercurrent diode effect (SDE) occurs in superconductors that lack inversion $\mathcal{I}$ and time-reversal \(\mathcal{T}\) symmetries (TRS), for example, due to the presence of spin-orbit coupling (SOC) and magnetic fields, respectively, see review Ref. \cite{NadeemFuhrerWang23} and references therein. Provided these symmetries are broken, critical supercurrents \(J_{c+}\) and \(J_{c-}\) flowing in opposite directions do not have to coincide in magnitude, namely \(J_{c+}\neq-J_{c-}\). Though the SDE has been proposed theoretically some time ago \cite{LevitovNazarovEliashberg85,Edelstein-JPCM96}, it has been observed for the first time only recently in layered heterostructures \cite{AndoYanaseOno20}. Nonreciprocal transport, namely magnetochiral anisotropy \cite{Rikken:PRL2001}, enhanced by superconducting fluctuations was observed in paraconductivity in a noncentrosymmetric monolayer transition metal dichalcogenide 1H-MoS\(_2\) \cite{WakatsukiNagaosa17}. It can be considered as a precursor effect as it requires breaking of the same symmetries. The related Josephson diode effect (JDE) and anomalous Josephson effect (AJE) in constrictions and microstructures predicted theoretically in Ref. \cite{HuWuDai07} and Refs. \cite{Josephson62,Fenton80,GeshkenbeinLarkin86,Yip95, KashiwayaTanaka00, RyazanovAarts01, KriveJohnson04,HouzetBuzdin07,BraudeNazarov07, Sigrist98, Buzdin05} respectively, also found their experimental confirmations in Ref. \cite{BaumgartnerManfraStrunk22} and Ref. \cite{SzombatiNadjPergeKouwenhoven16}, respectively.

The experimental discovery and subsequent observations of SDE \cite{LyuKwok21,BauriedlParadiso22,LinScheurerLi22,GutfreundAnahory23,HouMoodera23,SundareshRokhinson23,GengBergeretHeikkila23,PutilovBuzdin24,KealhoferBalentsStemmer23,SatchellBurnell23, YunKim23} spurred theoretical considerations of the microscopic origin of this effect \cite{WakatsukiNagaosa18,HoshinoNagaosa18,DaidoYanase22,DaidoYanase22_2,YuanFu22, HeNagaosa22,IlicBergeret22,KarabassovBobkovaVasenko22, IkedaDaidoYanase22, DaidoYanase23, ZhaiHe22, PutilovBuzdin23, ChazonoYanase23, Aoyama24, BanerjeeScheurer24, ZinklHamamotoSigrist22, KochanZutic23}. It was quickly realized that the original calculation of SDE in \cite{Edelstein-JPCM96} was incorrect as in the effective Ginzburg-Landau (GL) functional it only kept an anomalous term linear in the gradient (equivalently in the canonical momentum of the Cooper pair), i.e. the Lifshitz invariant \cite{Edelstein-JPCM96, MineevSamokhin08, BauerSigrist12, KochanZutic23}. While this term does break both \(\mathcal{I}\) and \(\mathcal{T}\) symmetries, in the absence of higher-order terms its effect on the critical current is nullified by the nonzero momentum \(q_0\) of the Cooper pairs in the ground state of the system, resulting in the so-called helical superconducting state \cite{MineevSamokhin94, Agterberg03, AgterbergKaur07, BauerSigrist12, SmidmanAgterberg17}, akin to the FFLO state \cite{FF,LO, SamokhinTruong17}. As a result, the ground state is symmetric under an accidental inversion-like symmetry \(\mathcal{I}': q+q_0\rightarrow -q+q_0\), which is enough to ensure that \(J_{c+}=-J_{c-}\) and that the SDE vanishes.

To account for this issue, theoretical works that followed the experimental observations of SDE included anomalous GL terms cubic in the canonical momentum \cite{HoshinoNagaosa18, DaidoYanase22, DaidoYanase22_2, YuanFu22, HeNagaosa22, IlicBergeret22, Yuan23, Yuan23II}. However, inconsistent results were found even in the simplest example of a helical superconductor: a two-dimensiona (2D) Rashba superconductor in an in-plane magnetic field with a pure \(s\)-wave pairing in the strong SOC limit. While most works found a superconducting diode coefficient \(\eta\stackrel{\text { def }}{=}(J_{c+}+J_{c-})/(J_{c+}-J_{c-})\propto \sqrt{T_c-T}\, H\) to the leading order \footnote{In the presence of either inversion or time reversal, \(J_{c-}=-J_{c+}\). Note that some authors use the opposite sign convention for \(J_{c-}\)}, where \(T_c\) is the critical temperature and \(H\) is the external in-plane magnetic field. In Ref. \cite{IlicBergeret22} it was found that \(\eta=0\) within the GL framework, and to linear order in \(H\), provided terms of fourth order in canonical momentum are included in the GL free energy.

We address the inconsistency in the literature by rederiving the GL-theory for the helical Rashba superconductor from a microscopic model. We confirm the result of Ref. \cite{IlicBergeret22} in a particular limit, and demonstrate that more generally  \(\eta\propto \sqrt{T_c-T}\, H^3\) to the leading order in \(H\).
The subtle reason behind this lies in the insufficiency of including terms of higher order in momentum to break the \(\mathcal{I}'\) symmetry to linear order in \(H\), as we prove. However, this symmetry is only approximate, as higher order terms in \(H\), symmetry-allowed deviations from the perfect 
\(s\)-wave pairing, and interband pairing all break it, thus resulting in a finite diodicity of superflow.

The rest of the paper is organized as follows. In Sec. \ref{SecGL} we review the general GL theory of SDE and determine the necessary order in canonical momentum needed to compute the superconducting diode coefficient and derive the formula for \(\eta\) in terms of the GL coefficients. In Sec. \ref{SecMicro}, we derive the GL coefficients for the particular case of the \(s\)-wave Rashba superconductor in an in-plane magnetic field for an arbitrary SOC strength. We then prove that \(\mathcal{I}'\) is an approximate symmetry in the strong SOC limit and show that this is a very special property of the \(s\)-wave Rashba superconductor. In particular, we compute leading corrections to \(\eta\) due to cubic terms in \(H\), as well as the correction linear in \(H\) due to interband pairing that is suppressed but non-zero in the limit of strong SOC. We discuss the ramifications of our findings for optimizing future superconducting diode designs in Sec. \ref{SecDisc}.


\begin{figure}
\includegraphics[width=0.45\textwidth]{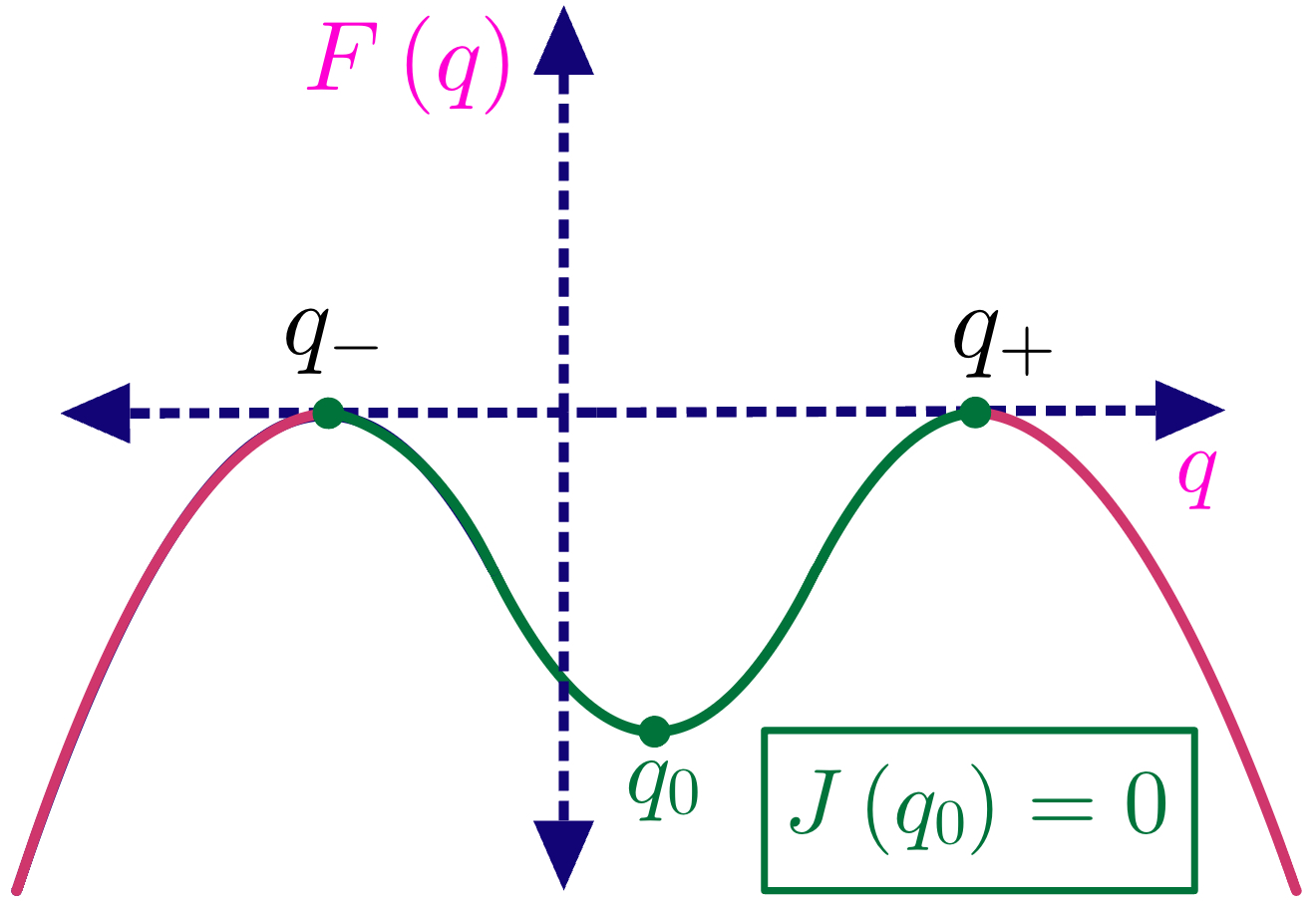} 
\caption{A schematic plot of the free energy $F\left(q\right)$ profile as a function of the collective momentum $q$ in the GL approximation near the critical temperature.}  \label{figII1}
\end{figure}

\section{Ginzburg-Landau Theory of SDE}\label{SecGL}

\begin{figure}
\includegraphics[width=0.45\textwidth]{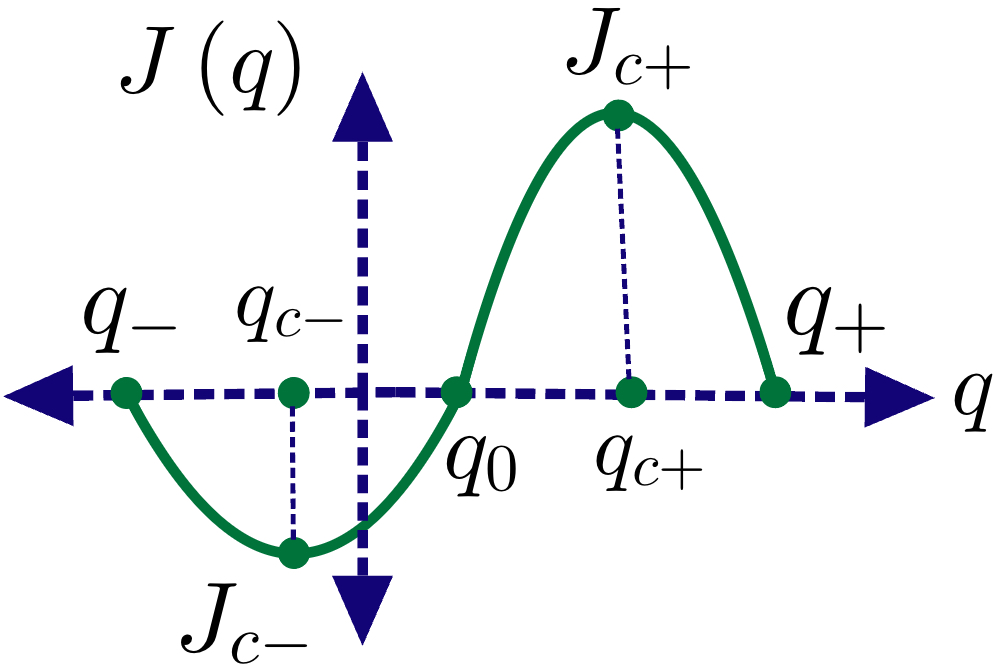} 
\caption{A schematic plot of a supercurrent corresponding to the free energy profile depicted in Fig. \ref{figII1}.}  \label{figII2}
\end{figure}

In this section we review the phenomenology of SDE in the Ginzburg-Landau framework. We start with the general Ginzburg-Landau expression for the free energy in the absence of \(\mathcal{I}\) and \(\mathcal{T}\), as considered in e.g. Refs. \cite{DaidoYanase22,YuanFu22,IlicBergeret22}:
\[\label{eq:II1}f(\Delta,q)=\alpha(q)\Delta^2+\beta(q)\Delta^4\]
with \(\alpha(q)=\sum_{n} \alpha_n q^n\) and \(\beta(q)=\sum_{n} \beta_n q^n\). This expression is obtained from the standard GL free energy with the ansatz \(\Delta(\mathbf{r})=\Delta e^{i\mathbf{q\cdot r}}\). We take \(\alpha_0=-A_0 t\) with \(t=(T_c-T)/T_c\) being the reduced temperature, where \(T_c\) is the critical temperature at zero magnetic field. The absence of \(\mathcal{I}\) and \(\mathcal{T}\) is reflected by the presence of the anomalous terms \(\alpha_{2n+1},\beta_{2n+1}\neq0\). The optimal order parameter is found to be \(\Delta(q)=\sqrt{-\alpha(q)/\beta(q)}\), with the condensation energy (expanding up to the fourth order in \(q\))
\[F(q)=-\frac{\alpha^2(q)}{4\beta(q)}\equiv-\frac{\tilde{\alpha}^2(q)}{4}\,,\]
with \(\tilde{\alpha}(q)=\sum_n\tilde{\alpha}_n q^n\) \footnote{We also assume $\beta(q)>0$ for all relevant $q$; otherwise higher-order terms in $\Delta$ need to be included.}. The supercurrent can then be obtained as \(J(q)=2\partial_q F(q)=-\tilde{\alpha}\partial_q\tilde{\alpha}/\beta_0\) \cite{DaidoYanase22}.
In the generic case, \(F(q)\) has two maxima at \(q_\pm\) corresponding to \(\tilde{\alpha}(q_\pm)=0\), at which \(F(q_\pm)=J(q_\pm)=0\), and no superconductivity exists for \(q>q_+\) and \(q<q_-\).
\(F(q)\) also has a nontrivial minimum with \(J(q_0)=0\) at some finite $q_0$ when \(\partial_q\tilde{\alpha}=0\) (see Fig \ref{figII1}). Between the minimum and the maxima of \(F(q)\), \(J(q)\) thus reaches a maximum and a minimum values \(J_{c+}\) and \(J_{c-}\), respectively, at some momenta \(q_{c\pm}\) determined by solving \(\partial_q J(q)=0\), or equivalently \(\tilde{\alpha}\partial_q^2 \tilde{\alpha}=-(\partial_q\tilde{\alpha})^2\) (see Fig \ref{figII2}).

With the accuracy up to the third order in the anomalous terms \(\tilde{\alpha}_{2n+1}\) one finds,
\[q_0=-\frac{\tilde{\alpha}_1}{2\tilde{\alpha}_2}+\frac{\tilde{\alpha}_1^2(2\tilde{\alpha}_1\tilde{\alpha}_4-3\tilde{\alpha}_2\tilde{\alpha}_3)}{8\tilde{\alpha}_2^4}\,.\]
In the presence of inversion symmetry or TRS the anomalous terms vanish and \(q_0=0\), as expected. When both are broken, the Cooper pairs carry momentum \(q_0\neq0\) in the ground state, which is therefore a helical superconductor \cite{Agterberg03,MineevSamokhin94,BauerSigrist12}. It is then natural to expand in powers of \(\delta q=q-q_0\), which yields
\[\tilde{\alpha}(q)=a_0+a_2\delta q^2+a_3\delta q^3+\dots\]
The phase transition occurs when \(a_0(T_c)=0\). Observe that if we keep terms only up to the quadratic order, \(F(q)\) (\(J(q)\)) is even (odd) under \(\mathcal{I}_{q_0}:\delta q\rightarrow -\delta q\), such that \(J_{c+}=-J_{c-}\) (and \(q_{c+}=-q_{c-}\)), and the SDE vanishes. To capture the SDE, it is therefore necessary (and sufficient) to work up to the cubic order in \(\delta q\), which close to the phase transition is restricted to be between \(\delta q_\pm\approx\pm\sqrt{-a_0/a_2}\). Note that close to the phase transition, \(q_\pm\rightarrow q_0\), such that the relevant range of \(\delta q\) shrinks, so the approximation becomes better closer to the phase transition.

Equipped with that approximation, it is possible to solve analytically for the critical supercurrents. The equation \(\tilde{\alpha}\partial_q^2 \tilde{\alpha}=-(\partial_q\tilde{\alpha})^2\) becomes, to the third order in \(\delta q\),
\[a_0a_2+3a_0a_3\delta q+3 a_2^2 \delta q^2+10 a_2 a_3\delta q^3=0\]
with two real solutions for \(\delta q\):
\[\delta q_{c\pm}=\pm\sqrt{\frac{-a_0}{3a_2}}+\frac{a_0a_3}{18a_2^2}.\]
Plugging this into the expression for the current, we find
\[J_{c\pm}=\frac{4a_0^2a_3}{9a_2}\pm\frac{4(-a_0)^{3/2}\sqrt{a_2}}{3\sqrt{3}},\]
and the superconducting diode coefficient is given by 
\[\eta=\frac{\sqrt{-a_0}a_3}{\sqrt{3}a_2^{3/2}}.\]
Note that these expressions are valid to any order in the anomalous terms. To the cubic order in the anomalous terms, we have
\begin{subequations}
\begin{align}
    &a_0=\tilde{\alpha}_0-\frac{\tilde{\alpha}^2_1}{4\tilde{\alpha}_2},\\
    &a_2 = \tilde{\alpha}_2 +\frac{3\tilde{\alpha}_1 (\tilde{\alpha}_1\tilde{\alpha}_4-\tilde{\alpha}_2\tilde{\alpha}_3)}{2\tilde{\alpha}^2_2}, \\
    &a_3 = \tilde{\alpha}_3 -\frac{2\tilde{\alpha}_1\tilde{\alpha}_4}{\tilde{\alpha}_2} \nonumber \\ &+\frac{\tilde{\alpha}_1^2(2\tilde{\alpha}_1\tilde{\alpha}_4^2+5\tilde{\alpha}_2^2\tilde{\alpha}_5-5\tilde{\alpha}_1\tilde{\alpha}_2\tilde{\alpha}_6-3\tilde{\alpha}_2\tilde{\alpha}_3\tilde{\alpha}_4)}{2\tilde{\alpha}_2^4}.
\end{align}
\end{subequations}
It is noteworthy that going to the third order in the anomalous terms requires going to the sixth order in \(\delta q\), while it is sufficient to go to the fourth order in \(\delta q\) if working to the first order in the anomalous terms, as found in Ref. \cite{IlicBergeret22} (in general, working to the \(n^{\text{th}}\) order in the anomalous terms requires working to the \((n+3)^{\text{rd}}\) order in \(\delta q\)). The coefficients \(\tilde{\alpha}_n\) can be obtained in terms of \(\alpha_n\) and \(\beta_n\), but the expressions are lengthy (it is sufficient to keep \(n\leq 6\)).
To the linear order in the anomalous terms and the leading order in \(a_0=\alpha_0/\sqrt{\beta_0}\), we have \(a_2=\alpha_2/\sqrt{\beta_0}\) and
\[a_3\approx\frac{2 \alpha_2 \alpha_3 \beta_0 - 4 \alpha_1 \alpha_4 \beta_0 - \alpha_2^2 \beta_1 + \alpha_1 \alpha_2 \beta_2}{2 \alpha_2 \beta_0^{3/2}},\]
while \(\beta_3\) and \(\beta_4\) do not enter the expression for \(\eta\):
\begin{equation}
\eta=\frac{2 \alpha_2 \alpha_3 \beta_0-4 \alpha_1 \alpha_4 \beta_0-\alpha_2^2 \beta_1+\alpha_1 \alpha_2 \beta_2}{2\sqrt{3} \alpha_2^{\frac{5}{2}}\beta_0}\sqrt{-\alpha_0}.\label{etaGL}
\end{equation}
One should observe that to this order in the anomalous terms, \(a_3=0\) if \(\tilde{\alpha}_2\tilde{\alpha}_3=2\tilde{\alpha}_1\tilde{\alpha}_4\) and the SDE vanishes; this was found to be the case for the Rashba superconductor considered in Ref. \cite{IlicBergeret22}.


\begin{figure}
\includegraphics[width=0.48\textwidth]{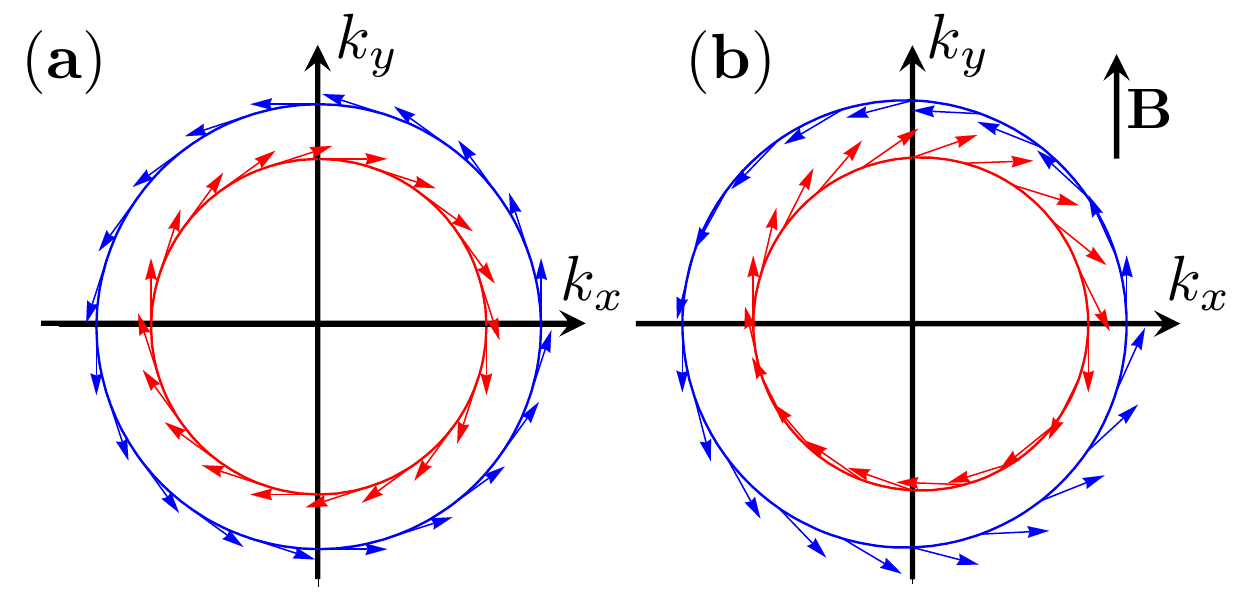} 
\caption{A schematic plot of the band structure and spin texture in the Rashba-Zeeman system with arrows denoting the spin orientation at the two spin split Fermi circles. (a) At zero Zeeman spin interaction, $\mathbf{B} = 0$ the two chiral bands are characterized by a winding $\pm1$ of the spin texture as one encircles the Fermi lines. (b) When a small $\mathbf{B}$ field is introduced the two
Fermi surfaces move toward each other in opposite directions perpendicular to the applied field. When Zeeman splitting
is relatively weak the chiral bands preserve a nontrivial winding, which becomes trivial in the opposite limit of the dominant
Zeeman spin splitting.}  \label{fig:HelicalBands}
\end{figure}

\section{Microscopic Calculation of GL Coefficients in the Helical Rashba Superconductor}\label{SecMicro}

\subsection{Model}\label{SubSecModel}

In this section we derive the coefficients of the GL theory for 2D noncentrosymmetric metals from microscopic theory by means of the Feynman diagram technique and, with the help of these coefficients investigate the influence of an external magnetic field on the critical current of the superconductor and consequently the superconducting diode coefficient $\eta$. The space inversion symmetry is broken by the presence of Rashba-type SOC term in the one-particle Hamiltonian \cite{Rashba1984} 
\begin{equation} \label{eq:III1}
H_{\mathrm{SO}}=\alpha_{R}([\mathbf{p} \times \mathbf{c}] \cdot \boldsymbol{\sigma}),
\end{equation}
where $\mathbf{p}$ is the particle momentum,  $\mathbf{c}$ is the unit vector along the direction of the asymmetric potential gradient perpendicular to the 2D metal, $\boldsymbol{\sigma}$ is the Pauli spin matrix-vector, and the parameter $\alpha_{R}$ denotes the strength of the spin-orbit interaction, which has units of velocity. Here and hereafter, we use natural units in which $\hbar$ and Boltzmann's constant $k_B$ are set to unity. By lifting the spin degeneracy of the conduction electrons, the SOC forms two energy branches with positive and negative helicities (the projection of the spin of an electron with momentum $\mathbf{p}$ on the direction $\mathbf{c} \times \mathbf{p}$) with energies which, on the assumption of the isotropic electron mass, are $\epsilon_{ \pm}(p)=$ $\frac{p^2}{2 m} \pm \alpha_{R} p$. So, the states of positive and negative helicity acquire different energies, see Fig. \ref{fig:HelicalBands} for the illustration. 

We emphasize that the SOC constant $\alpha_{R}$ enters the problem through two parameters in the clean limit: 
\begin{equation}
\delta=\frac{m\alpha_{R}}{p_F} \quad \text{and} \quad \kappa=\frac{\alpha_{R} p_F}{\pi T_c},
\end{equation}
where $m$ is the electron effective mass, $p_F$ is the Fermi momentum in the absence of SOC. The parameter $\delta$ is assumed to be small so in equations all powers of $\delta$ in excess of the first have been ignored. But the parameter $\kappa$ does not limit the theory and is allowed below to take any value in contrast to the earlier theoretical works involving the SDE \cite{HeNagaosa22,IlicBergeret22} where the limit of $\kappa \to \infty$ was taken from the outset. We closely follow Edelstein's approach \cite{Edelstein-JPCM96,edelstein_ginzburg-landau_2021} in calculating the GL coefficients from the microscopic perspective in part because this approach allows us to analyze the SDE for arbitrary values of the parameter \(\kappa\) and it is free from the additional approximations of a semiclassical method.


\begin{figure}
\includegraphics[width=0.48\textwidth]{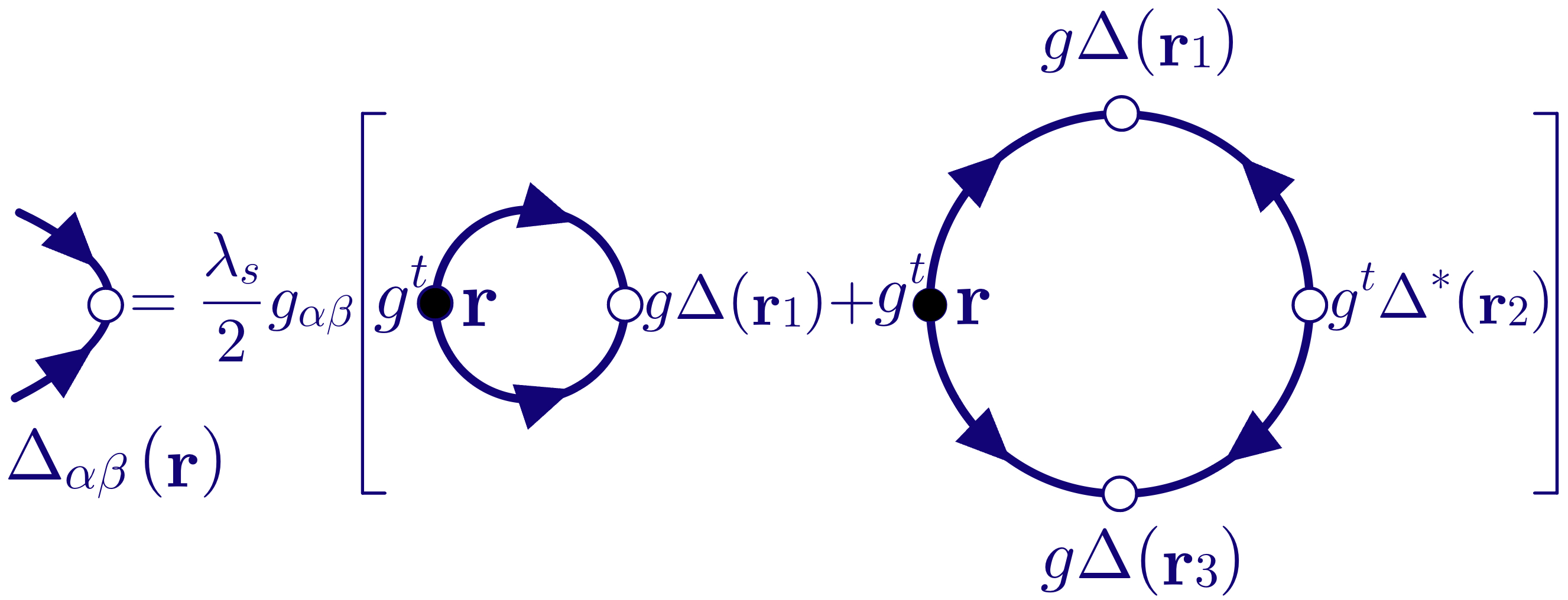} 
\caption{The diagrammatic equation for $\Delta_{\alpha \beta}(\mathbf{r})$. Here $\lambda_s$ is the pairing constant, a thick clockwise fermion line denotes $G^{(11)}\left(i \omega_n|\mathbf{r}_1, \mathbf{r}_2\right)$, while an anticlockwise line denotes $G^{(22)}\left(i \omega_n|\mathbf{r}_1, \mathbf{r}_2\right)$ defined by Eqs. (\ref{eq:IIIA1}) and (\ref{eq:IIIA2}) and the superscript $t$ denotes matrix transposition.}  \label{fig1}
\end{figure}

\subsection{Diagram technique and calculation}\label{SubSecFormulation}

The starting point of our analysis is the usual BCS self-consistency equation for the order parameter, $\Delta_{\gamma \rho}(\mathbf{r})=g_{\gamma \rho} \Delta(\mathbf{r})$, $g=i \sigma_y$. Here we assumed an $s$-wave pairing. For the purposes of deriving an effective GL functional we expand the self-consistency equation in powers of $\Delta(\mathbf{r})$. This expansion procedure is depicted diagrammatically on Fig. \ref{fig1}. The building blocks of the diagrammatic series include matrix Green's function 
\begin{equation}\label{eq:IIIA1}
\begin{aligned}
&\widehat{G}_{\alpha \beta}\left(i \omega_n|\mathbf{r}_1, \mathbf{r}_2\right)\\&=\begin{pmatrix}
G^{(11)}\left(i \omega_n | \mathbf{r}_1, \mathbf{r}_2\right) & 0 \\
0 & G^{(22)}\left(i \omega_n | \mathbf{r}_1, \mathbf{r}_2\right)
\end{pmatrix}_{\alpha \beta},
\end{aligned}
\end{equation}
which is depicted by thick lines and represents the system in the normal state subject to the applied magnetic field. It obeys the following Dyson equation:
\begin{equation}\label{eq:IIIA2}
\begin{aligned}
\widehat{G}\left(i \omega_n | \mathbf{r}_1, \mathbf{r}_2\right)= & \widehat{G}^{(0)}\left(i \omega_n | \mathbf{r}_1-\mathbf{r}_2\right)+\int_{\mathbf{r}} \widehat{G}^{(0)}\left(i \omega_n | \mathbf{r}_1-\mathbf{r}\right) \\
& \times\widehat{V}(\mathbf{r}) \widehat{G}\left(i \omega_n | \mathbf{r}, \mathbf{r}_2\right).
\end{aligned}
\end{equation}
Here $\widehat{G}^{(0)}\left(i \omega_n|\mathbf{r}_1-\mathbf{r}_2\right)$ is the Green's matrix function of the clean system without the external field, which can be found from the equation of motion
\begin{equation}\label{eq:IIIA3}
\begin{aligned}
& \left(\begin{array}{cc}
i \omega_n-H\left(-i \boldsymbol{\nabla}_1\right) & 0 \\
0 & i \omega_n-H^t\left(i \boldsymbol{\nabla}_1\right)
\end{array}\right)_{\beta \kappa} \\
& \times \widehat{G}_{\kappa \rho}^{(0)}\left(i \omega_n | \mathbf{r}_1-\mathbf{r}_2\right)=\left(\begin{array}{cc}
1 & 0 \\
0 & 1
\end{array}\right) \delta\left(\mathbf{r}_1-\mathbf{r}_2\right) \delta_{\beta \rho},
\end{aligned}
\end{equation}
with the Hamiltonian
\begin{equation}\label{eq:IIIA4}
H_{\beta \gamma}(-i \boldsymbol{\nabla})=-\frac{\nabla^2}{2 m} \delta_{\beta \gamma}+\alpha_{R}(-i \boldsymbol{\nabla} \times \mathbf{c}) \cdot \boldsymbol{\sigma}_{\beta \gamma}.
\end{equation}
Here $\omega_n=\pi T(2 n+1)$ with $n\in\mathbb{Z}$ are fermion Matsubara frequencies and the superscript $t$ denotes matrix transposition. For simplicity, we have assumed that the spectrum of the electrons in the absence of $H_{\text{SO}}$ and the interparticle interaction is isotropic. 

\begin{figure}[t!]
\includegraphics[width=0.48\textwidth]{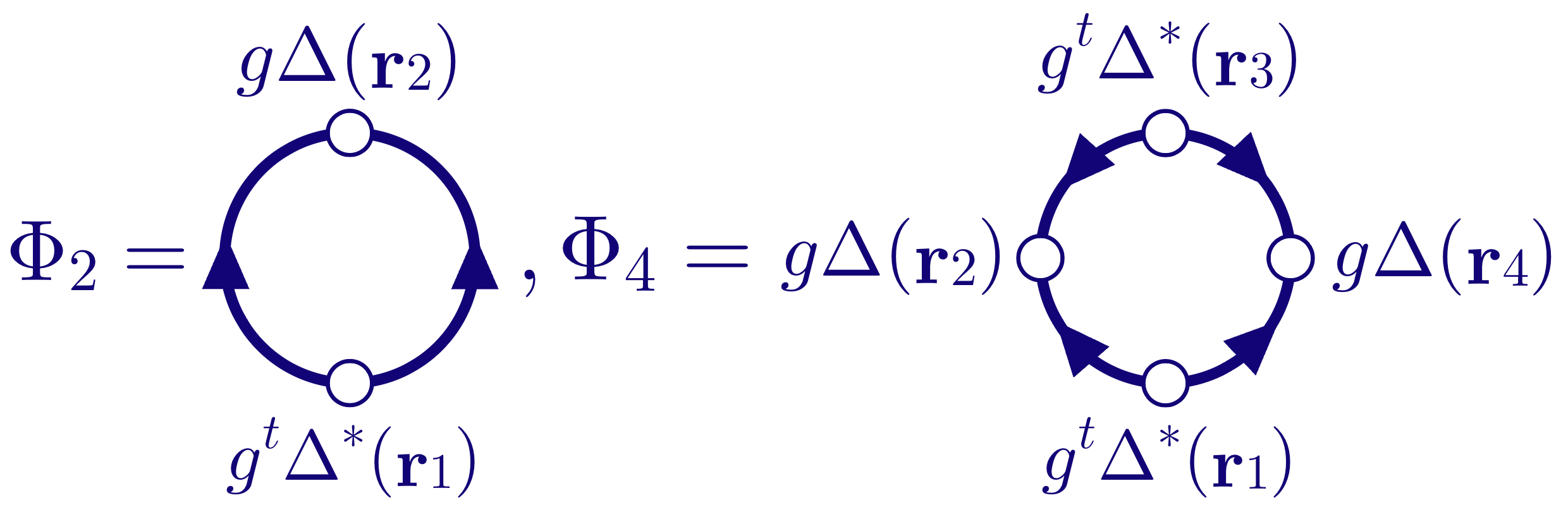} 
\caption{Loop diagrams for the GL functional.}  \label{fig2}
\end{figure}

The convolution term in the Dyson equation (\ref{eq:IIIA2}) contains a matrix 
\begin{equation}\label{eq:IIIA5}
\widehat{V}_{\alpha \beta}(\mathbf{r})=\left(\begin{matrix}
V_{\alpha \beta}(\mathbf{r}) & 0 \\
0 & -V_{\alpha \beta}^{t}(\mathbf{r})
\end{matrix}\right)
\end{equation}
that takes into account the effect of the magnetic field $\mathbf{B}(\mathbf{r})$,
\begin{equation}\label{eq:IIIA6}
V_{\gamma \rho}(\mathbf{r})=\mu_B \boldsymbol{\sigma}_{\gamma \rho} \cdot \mathbf{B}(\mathbf{r}),
\end{equation}
where $\mu_B$ is the Bohr magneton. 

The equation for $\hat{\Delta}$ obtained in Fig. \ref{fig1} corresponds to a minimum of thermodynamic potential 
\begin{equation}\label{eq:IIIA7}
\frac{\delta \Omega}{\delta \Delta^*(\mathbf{r})}=0
\end{equation}
which takes the usual form 
\begin{equation}\label{eq:IIIA8}
\Omega=\frac{1}{\lambda_s} \int_{\mathbf{r}}|\Delta(\mathbf{r})|^2+\frac{1}{2}\Phi_2+\frac{1}{4} \Phi_4.
\end{equation}
Here $\lambda_s$ is the pairing constant, $\Phi_2$ and $\Phi_4$ are quadratic and quartic functionals of $\hat{\Delta}(\mathbf{r})$, respectively, defined diagrammatically in Fig. \ref{fig2}. The next technical task is to expand the diagrams of Fig. \ref{fig2} into a series in the small external field $\mathbf{B}(\mathbf{r})$.

\begin{figure}
\includegraphics[width=0.48\textwidth]{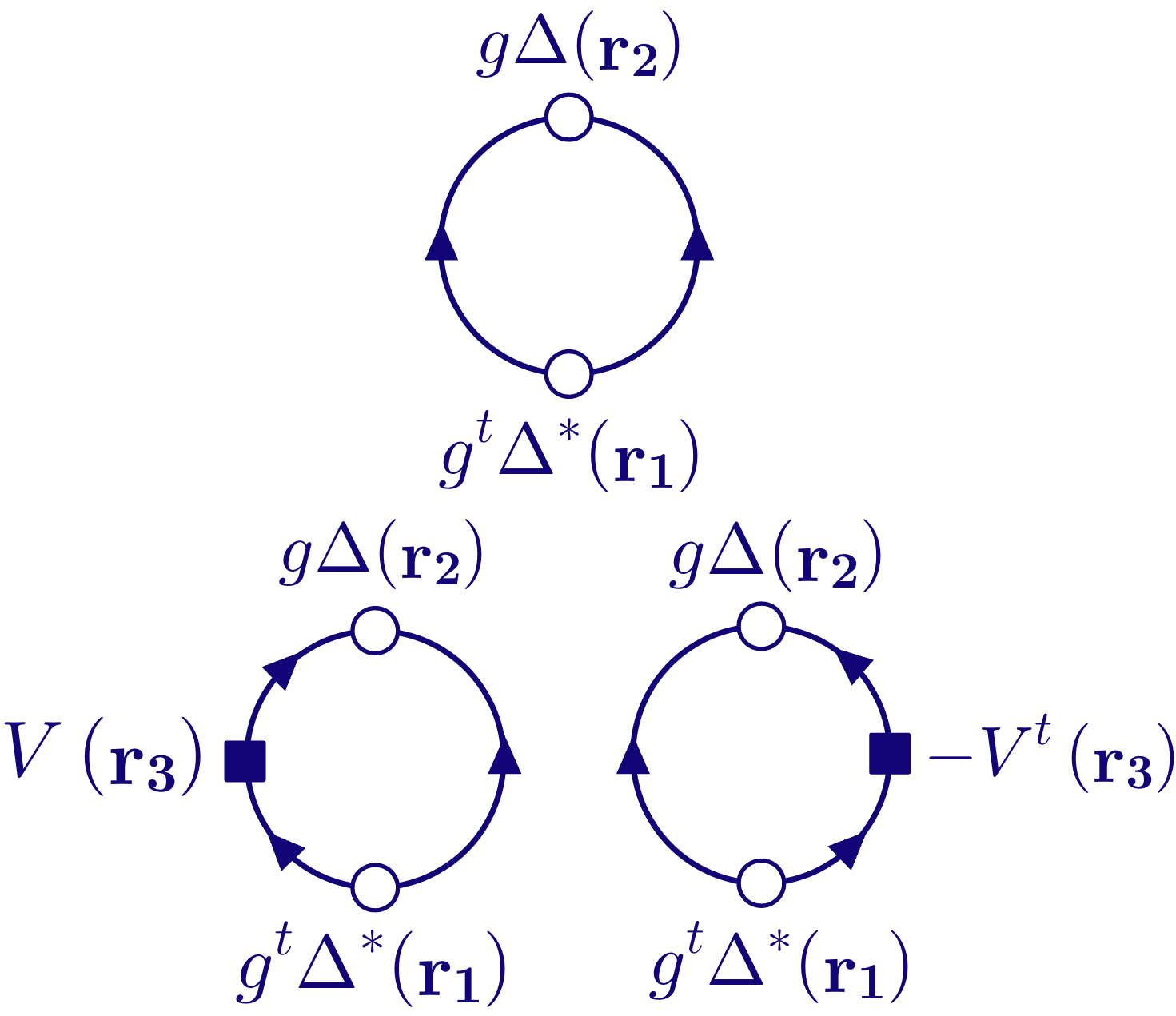} 
\caption{Diagrammatic representation for $\Phi_{\mathrm{2.c}}$ (top row) and $\Phi_{\mathrm{2.an}}$ (bottom row). The solid square denotes $V(\mathbf{r})$ if it is placed on a clockwise solid line, whereas being placed on an anticlockwise line it denotes $-V^{t}(\mathbf{r})$.}  \label{fig3}
\end{figure}

As a result, the functional $\Phi_2$ transforms into a sum of the conventional term $\Phi_\mathrm{2.c}$ and the anomalous term $\Phi_{\mathrm{2.an}}$ presented by diagrams in Fig. \ref{fig3}. Unlike Fig. \ref{fig2}, thin solid lines in Fig. \ref{fig3} correspond to the normal system but without the magnetic field term in the Green's function. The solid square denotes $V(\mathbf{r})$ if it is placed on a clockwise solid line, whereas being placed on an anticlockwise line it denotes $-V^{t}(\mathbf{r})$.

Diagrams corresponding to the functional $\Phi_4$ can be represented in exactly the same way. The conventional term $\Phi_\mathrm{4.c}$ can be obtained by replacing the thick solid lines in Fig. \ref{fig2} by thin solid lines and the anomalous term  $\Phi_\mathrm{4.an}$ can be obtained from $\Phi_\mathrm{4.c}$ by attaching one $V^{}(\mathbf{r})$ or one $-V^{t}(\mathbf{r})$ solid square in any one of the four thin solid lines of the $\Phi_\mathrm{4.c}$ diagram, depending on whether the thin solid line is clockwise or anticlockwise respectively. 

\begin{figure}
\includegraphics[width=0.48\textwidth]{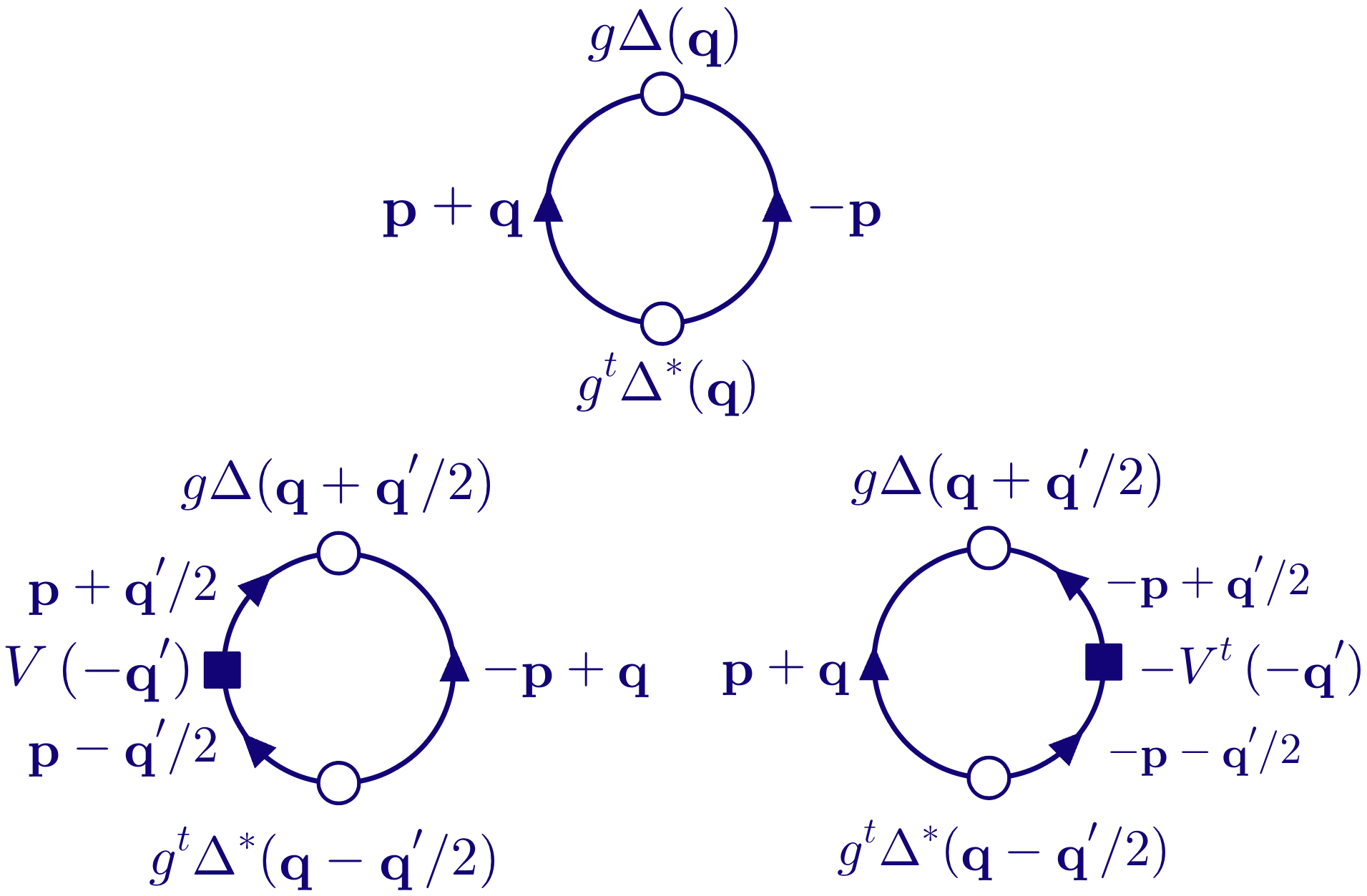} 
\caption{Diagrams of the Fig. \ref{fig3} in the momentum space representation.}  \label{fig4}
\end{figure}

The relation between the momentum and coordinate representations in the presence of the SOC has the form
\begin{subequations}\label{eq:IIIA9}
\begin{align}
&\widehat{G}_{\kappa \rho}^{(0)}\left(i \omega_n, \mathbf{r}-\mathbf{r}^{\prime}\right)=\nonumber \\
&\int_{\mathbf{p}} e^{i \mathbf{p} \cdot\left(\mathbf{r}-\mathbf{r}^{\prime}\right)}\left(\begin{matrix}
G\left(i \omega_n, \mathbf{p}\right) & 0 \\
0 & -G^t\left(-i \omega_n,-\mathbf{p}\right)
\end{matrix}\right)_{\kappa \rho},\\
&G_{\kappa \rho}\left(i \omega_n, \mathbf{p}\right)  =\sum_{\nu= \pm} \Pi_{\kappa \rho}^{(\nu)}(\mathbf{p}) G_{(\nu)}\left(i \omega_n, p\right), \\
&\Pi^{( \pm)}(\mathbf{p})  =\frac{1}{2}\left(\sigma_0 \pm \frac{(\mathbf{p} \times \mathbf{c}) \cdot \boldsymbol{\sigma}}{|\mathbf{p} \times \mathbf{c}|}\right).
\end{align}
\end{subequations}
Here $\int_{\mathbf{p}}=\int \frac{d^2 p}{(2 \pi)^2}$, $\Pi^{( \pm)}(\mathbf{p})$ is the operator of projection onto states with a definite helicity, 
\begin{equation}\label{eq:IIIA10}
G_{(\nu)}\left(i \omega_n, p\right)=\left[i \omega_n-\xi_{(\nu)}(p)\right]^{-1},
\end{equation}
and $\xi_{( \pm)}(p)=\epsilon_{ \pm}(p)-\mu$. This Green function in Eq. (\ref{eq:IIIA9}) is the basic tool for subsequent work. The only significant difference between the diagram technique here and the standard one is in the spinor structure of the Green function and the changed form of the velocity operator,
\begin{equation}\label{eq:IIIA11}
\mathbf{v}_{\beta \gamma}(\mathbf{p})=\mathrm{i}\left[H_{\beta \gamma}(\mathbf{p}), \mathbf{r}\right]=\frac{\mathbf{p}}{m} \delta_{\beta \gamma}+\alpha_{R}(\mathbf{c} \times \boldsymbol{\sigma})_{\beta \gamma}
\end{equation}
which, along with the usual scalar part, also has a spin component. For the following, it is convenient to also introduce the reversed Green's function, $G_{\kappa \rho}^{(\text{r})}\left(i \omega_n, \mathbf{p}\right)$ via the equation
\begin{equation}\label{eq:IIIA12}
-G_{\kappa \rho}^t\left(-i \omega_n,-\mathbf{p}\right)=g_{\kappa \gamma}^t G_{\gamma \beta}^{(\text {r})}\left(i \omega_n, \mathbf{p}\right) g_{\beta \rho} .
\end{equation}
Then
\begin{subequations}\label{eq:IIIA13}
\begin{align}
G_{\kappa \rho}^{(\mathrm{r})}\left(i \omega_n, \mathbf{p}\right) & =\sum_{\nu= \pm} \Pi_{\kappa \rho}^{(\nu)}(\mathbf{p}) G_{(\nu)}^{(\mathrm{r})}\left(i \omega_n, p\right), \\
G_{(\nu)}^{(\mathrm{r})}\left(i \omega_n, p\right) & =\left[i \omega_n+\xi_{(\nu)}(p)\right]^{-1}.
\end{align}
\end{subequations}
In deriving Eqs. (\ref{eq:IIIA13}), use has been made of the equality
\[\label{eq:IIIA14}
g \Pi^{t( \pm)}(-\mathbf{p}) g^t=\Pi^{( \pm)}(\mathbf{p}),
\]
which is a consequence of the readily verifiable identity $g \sigma^t g^t=$ $-\sigma$. Eq. (\ref{eq:IIIA6}) in momentum space take the form
\[\label{eq:IIIA15}
V_{\beta \gamma}(\mathbf{r}) \rightarrow V_{\beta \gamma}(\mathbf{q})=\mu_B \boldsymbol{\sigma}_{\beta \gamma} \cdot \mathbf{B}(\mathbf{q})
\]
As we can observe in the momentum representation, clockwise-directed fermion lines of a diagram become $G_{\beta \gamma}\left(i \omega_n, \mathbf{p}\right)$ while anticlockwise-directed lines become $-G_{\rho \zeta}^t\left(-i \omega_n,-\mathbf{p}\right)$. Thus the diagrams of Fig. \ref{fig3} are converted to the diagrams in Fig. \ref{fig4} in the momentum representation.

\begin{figure}
\includegraphics[width=0.48\textwidth]{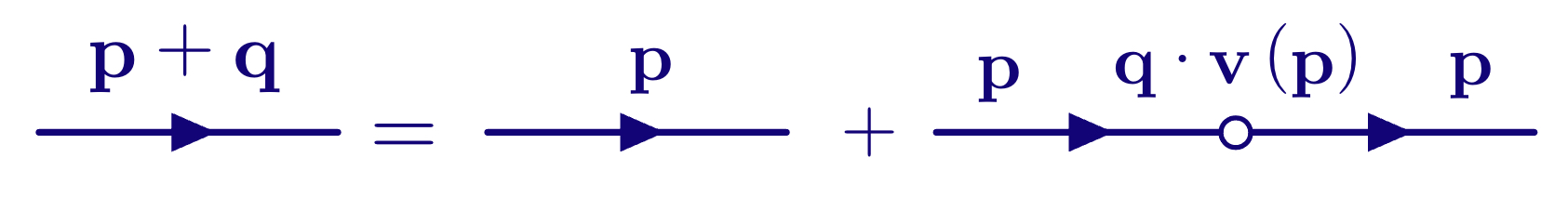} 
\caption{Diagrammatic representation of Eq. (\ref{eq:IIIA16}).}  \label{fig5}
\end{figure}

There is yet one more perturbation energy which has to be taken into account, that is the Doppler energy $H_D(\mathbf{p}, \mathbf{q})=\frac{1}{2} \mathbf{q} \cdot \mathbf{v}(\mathbf{p})$, acquired by the electron of momentum $\mathbf{p}$ by interaction with the external field (or with the order parameter) bearing momentum $\mathbf{q}$. Since all the Bose-type fields $(\mathbf{B}$ and $\hat{\Delta})$ are assumed to be slowly varying over space, the momenta with which they enter the diagrams are much smaller than both $p_F$ and $\xi_0^{-1}$. Therefore, these two energies can be considered as small perturbations. So, the Green's functions can be safely expanded into a series of small parameters $q / p_F$ and $q^{\prime}/ p_F$ as
\begin{equation}\label{eq:IIIA16}
\begin{aligned}
& G^{(0)}\left(i \epsilon_n, \mathbf{p}+\mathbf{q}\right) \\
& =G^{(0)}\left(i \epsilon_n, \mathbf{p}\right)+G^{(0)}\left(i \epsilon_n, \mathbf{p}\right)[\mathbf{q} \cdot \mathbf{v}(\mathbf{p})] G^{(0)}\left(i \epsilon_n, \mathbf{p}\right)+\cdots
\end{aligned}
\end{equation}
or in the diagram language as shown in Fig. \ref{fig5}.

\begin{figure}
\includegraphics[width=0.48\textwidth]{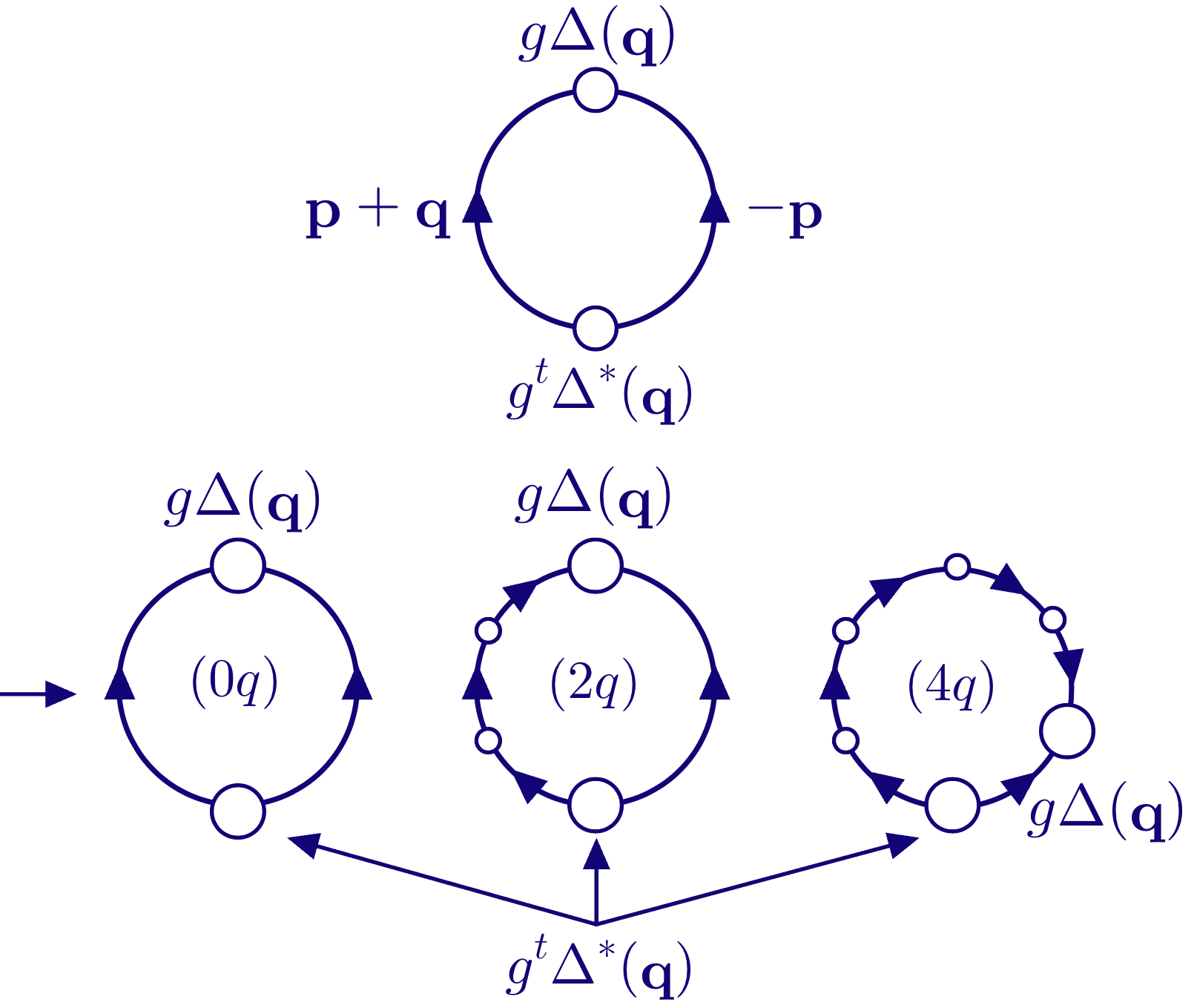} 
\caption{The contribution corresponding to $\Phi_{\mathrm{2.c}}$ to zeroth order in the magnetic field. Here, all the clockwise lines have the momenta $\mathbf{p}$ and all the anticlockwise lines have the momenta $-\mathbf{p}$. The small circles all represent $\mathbf{q} \cdot \mathbf{v}(\mathbf{p})$ if they appear on a clockwise line, $-\mathbf{q} \cdot \mathbf{v}^{\mathbf{t}}(-\mathbf{p})$ if they appear on an anticlockwise line.}  \label{fig6}
\end{figure}

\begin{figure}
\includegraphics[width=0.34\textwidth]{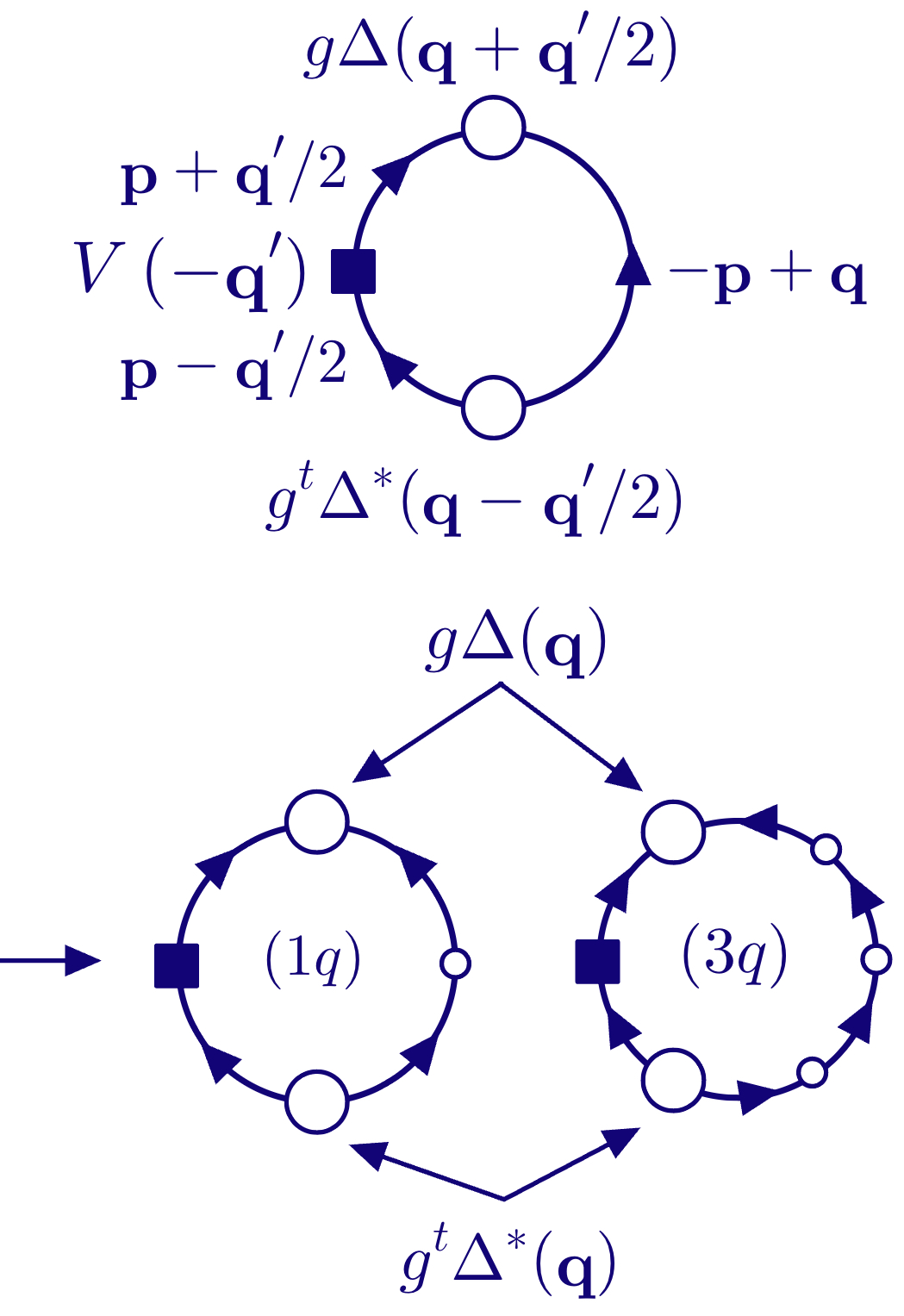} 
\caption{The contribution of the first diagram corresponding to $\Phi_{\mathrm{2.an}}$ to first order in the magnetic field.}  \label{fig7}
\end{figure}

\begin{figure}
\includegraphics[width=0.34\textwidth]{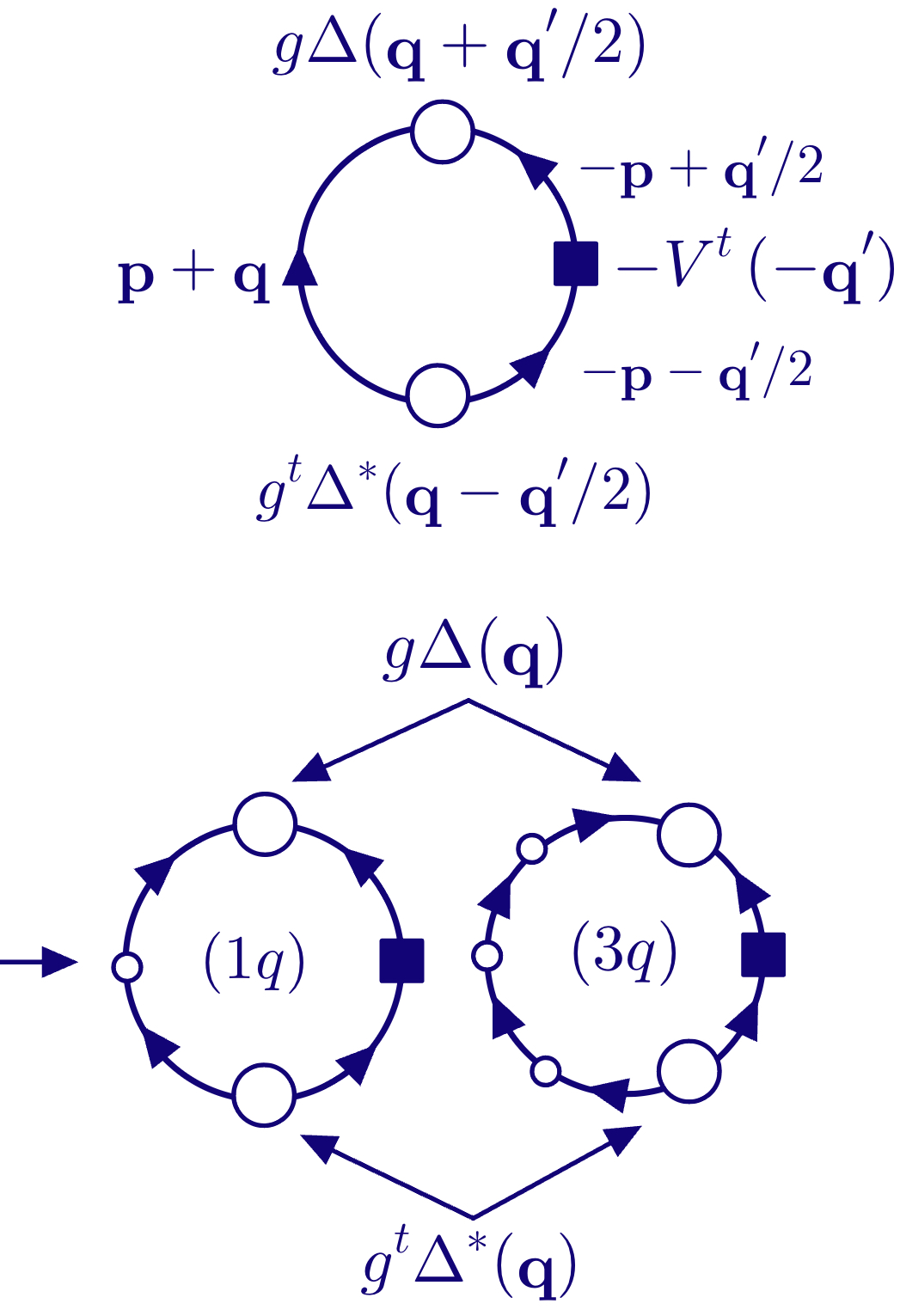} 
\caption{The contribution of the second diagram corresponding to $\Phi_{\mathrm{2.an}}$ to first order in the magnetic field.}  \label{fig8}
\end{figure}

Thus the $q-$expansion of the diagrams in Fig. \ref{fig4} corresponding to $\Phi_\mathrm{2.c}$ and $\Phi_\mathrm{2.an}$ can be carried out. We can also neglect the $q^{\prime}$ dependence of the magnetic field considering small applied magnetic field. From the GL analysis, the only pertinent diagrams for the superconducting diode coefficient $\eta$ to linear order in magnetic field are obtained by the $q-$expansion of the diagrams upto $\order{q^4}$. Because the diagrams odd(even) in $q$ in the $q-$expansion of $\Phi_\mathrm{2.c}$ ($\Phi_\mathrm{2.an}$) can be shown to vanish, the nonvanishing diagrams corresponding to $\Phi_\mathrm{2.c}$ and $\Phi_\mathrm{2.an}$ upto $\order{q^4}$ are depicted in Figs. \ref{fig6}-\ref{fig8}.

\begin{figure}
\includegraphics[width=0.48\textwidth]{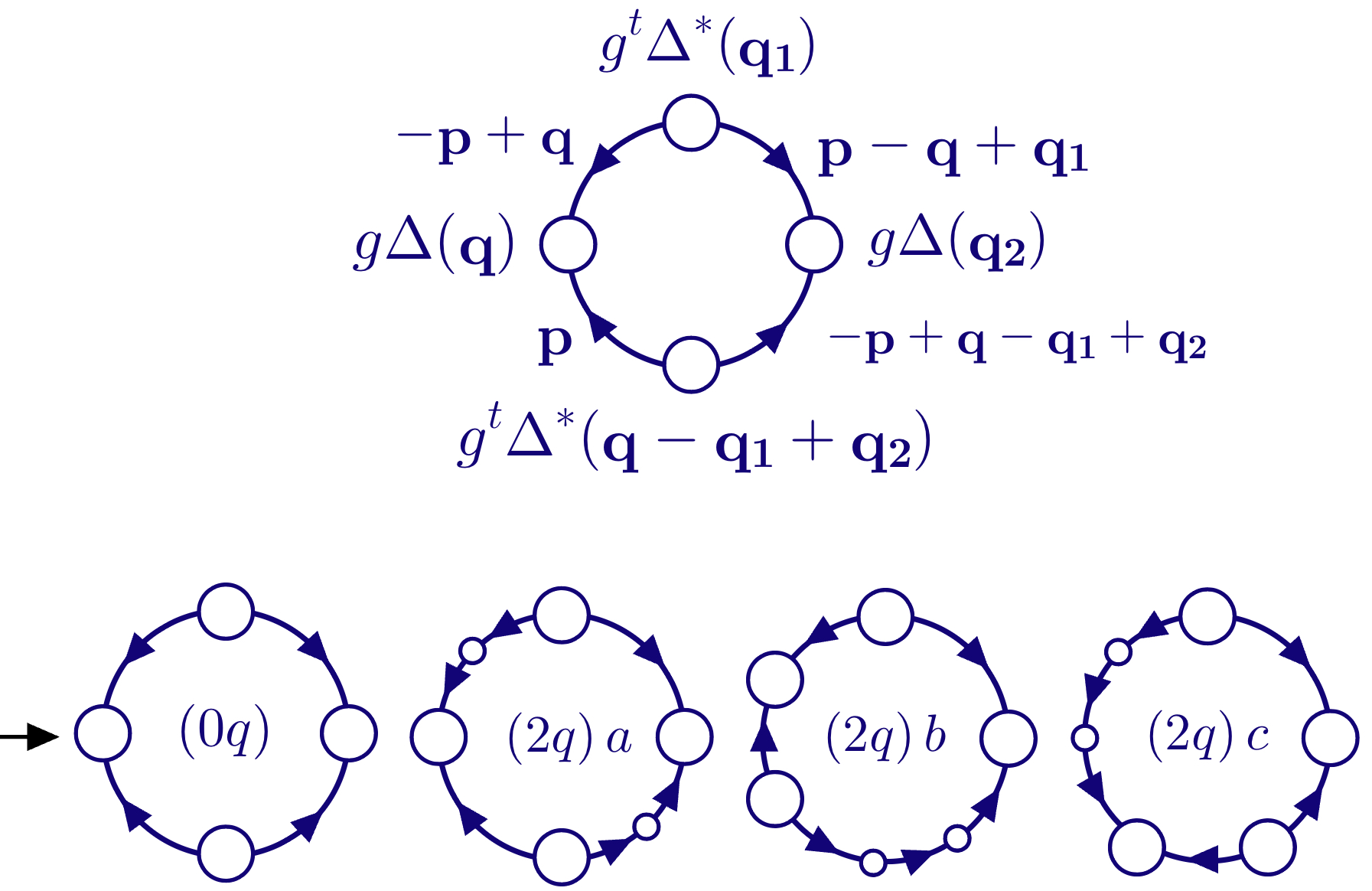} 
\caption{The contribution corresponding to $\Phi_{\mathrm{4.c}}$ to zeroth order in the magnetic field upto $\order{q^4}$. Here, we have assumed $\mathbf{q}=\mathbf{q_1}=\mathbf{q_2}$.}  \label{fig9}
\end{figure}

\begin{figure}
\includegraphics[width=0.46\textwidth]{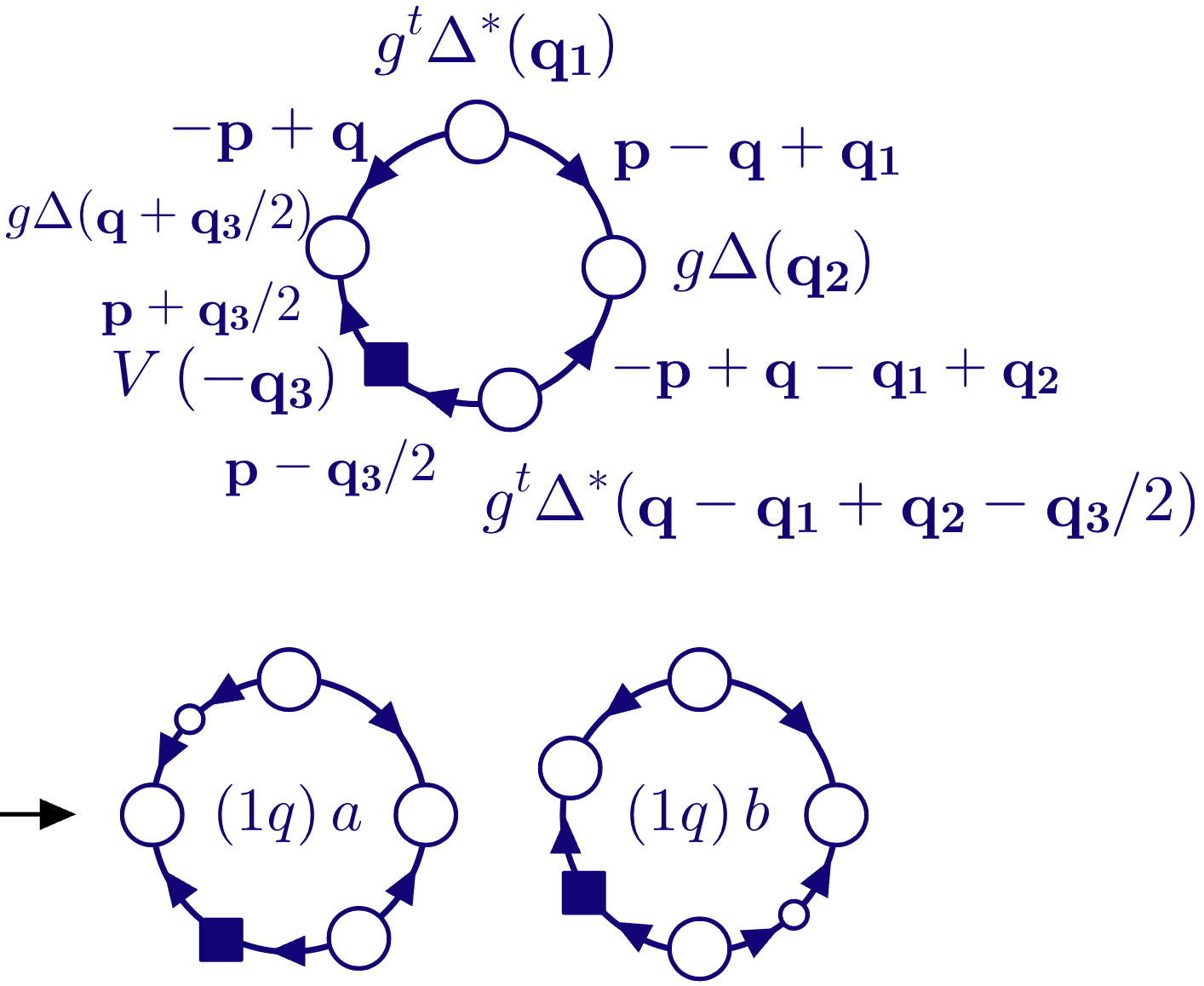} 
\caption{The contribution corresponding to $\Phi_{\mathrm{4.an}}$ to first order in the magnetic field upto $\order{q}$. Here, we have assumed $\mathbf{q}=\mathbf{q_1}=\mathbf{q_2}$ and $V$ is independent of $\mathbf{q_3}$.}  \label{fig10}
\end{figure}

The $q-$expansion of the diagrams corresponding to $\Phi_\mathrm{4.c}$ and $\Phi_\mathrm{4.an}$ can be obtained similarly and they are shown in Figs. \ref{fig9} and \ref{fig10}. Note that in Fig. \ref{fig10} we have attached the $V$ solid square to only one of the four thin solid lines of the diagrams corresponding to $\Phi_\mathrm{4.an}$. Four such analogous set of diagrams obtained by attaching the $V$ solid square to any of the other three thin solid lines of the diagrams corresponding to $\Phi_\mathrm{4.an}$ can be shown to each contribute equally to $\Phi_\mathrm{4.an}$.\\

Calculating the diagrams using Edelstein's approach (see Appendices A, B and C) and comparing Eq. (\ref{eq:II1}) with Eq. (\ref{eq:IIIA8}) yields the following for the coefficients of $\alpha(\mathbf{q})$:
\begin{subequations}
\begin{equation}
\alpha_0=-\nu\left(\frac{T_c-T}{T_c}\right),
\end{equation}
\begin{equation}
\alpha_1=-\frac{\nu\alpha_{R}}{2} (\left[\mathbf{h} \times \hat{\mathbf{q}}\right] \cdot \mathbf{c}) \sum_{\omega_n>0} \frac{2\pi T\left(\alpha_{R} p_F\right)^2}{ \omega_n^3\left[ \omega_n^2+\left(\alpha_{R} p_F\right)^2\right]},
\end{equation}
\begin{equation}
\alpha_2=\frac{\nu v_F^2}{8} \sum_{\omega_n>0} \frac{2\pi T}{\omega_n^3},
\end{equation}
\begin{equation}
\begin{split}
&\alpha_3=\frac{\nu\alpha_{R}v_F^2}{32} (\left[\mathbf{h} \times \hat{\mathbf{q}}\right] \cdot \mathbf{c})\times \; \\&2\pi T \sum_{\omega_n>0} \frac{15 \omega_n^4 \left(\alpha_{R} p_F\right)^2+17\omega_n^2 \left(\alpha_{R} p_F\right)^4+6\left(\alpha_{R} p_F\right)^6}{ \omega_n^5\left[ \omega_n^2+\left(\alpha_{R} p_F\right)^2\right]^3},
\end{split}
\end{equation}
\begin{equation}
\alpha_4=-\frac{3\nu v_F^4}{128} \sum_{\omega_n>0} \frac{2\pi T}{ \omega_n^5},
\end{equation}
\end{subequations}
and the coefficients of $\beta(\mathbf{q})$ are as follows:
\begin{subequations}
\begin{equation}
\beta_0=\frac{\nu}{4} \sum_{\omega_n>0} \frac{2\pi T}{ \omega_n^3},
\end{equation}
\begin{align}
&\beta_1=\frac{\nu\alpha_{R}}{4}(\left[\mathbf{h} \times \hat{\mathbf{q}}\right] \cdot \mathbf{c})\nonumber \times \\ 
&2\pi T \sum_{\omega_n>0} \frac{3\left(\alpha_{R} p_F\right)^4+5\left(\alpha_{R} p_F\right)^2\omega_n^2}{ \omega_n  ^5\left[ \omega_n ^2+\left(\alpha_{R} p_F\right)^2\right]^2},
\end{align}
\begin{equation}
\beta_2=-\frac{3\nu v_F^2 }{16} \sum_{\omega_n>0} \frac{2\pi T}{ \omega_n^5},
\end{equation}
\begin{equation}
\begin{split}
&\beta_3=-\frac{\nu\alpha_{R}v_F^2 }{64}(\left[\mathbf{h} \times \hat{\mathbf{q}}\right] \cdot \mathbf{c}) \sum_{\omega_n>0} \frac{2\pi T\left(\alpha_{R} p_F\right)^2}{ \omega_n^7\left[ \omega_n ^2+\left(\alpha_{R} p_F\right)^2\right]^4}\times
\\&[45(\alpha_{R} p_F)^6+176\omega_n^2 (\alpha_{R} p_F)^4+253\omega_n^4 (\alpha_{R} p_F)^2+154 \omega_n^6],
\end{split}
\end{equation}
\begin{equation}
\beta_4=\frac{45\nu v_F^4}{512} \sum_{\omega_n>0} \frac{2\pi T}{\omega_n^7},
\end{equation}
\end{subequations}
where $\nu=\frac{m}{2\pi}$ is the density of states of a simple parabolic band in 2D, $\mathbf{h}=\mu_B \mathbf{B}$, and $\hat{\mathbf{q}}$ is the unit vector along $\mathbf{q}$. 

\begin{figure}\label{fig:f-kappa}
\includegraphics[width=0.48\textwidth]{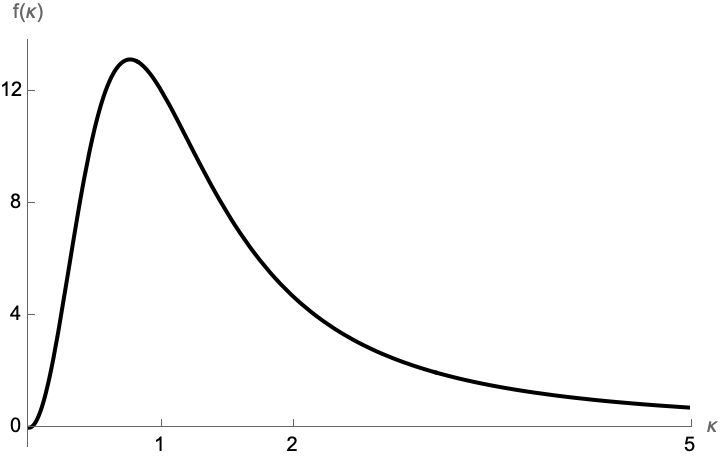} 
\caption{Numerical plot of $f\left(\kappa\right)$ as defined in Eq. \eqref{eq:eta-f}.} 
\end{figure} 

As the final step, we can compute the Matsubara sums above (see Appendix D) and derive an expression for the SDE coefficient $\eta$. This calculation can be carried out for the arbitrary value of $\kappa$. We find the following result 
\begin{subequations}
\begin{equation}\label{eq:eta}
\eta=\frac{\delta}{7\sqrt{42\zeta^3(3)}} \left(\frac{\left[\mathbf{h} \times \hat{\mathbf{q}}\right] \cdot \mathbf{c}}{\pi T_c}\right) \sqrt{\frac{T_c-T}{T_c}}f\left(\kappa\right),
\end{equation}
where 
\begin{align}\label{eq:eta-f}
&f\left(\kappa\right)=\frac{1}{\kappa^4}\left[14\kappa^2\zeta\left(3\right)+16\operatorname{Re}\left(\psi\left(\frac{1}{2}+\frac{i \kappa}{2}\right)-\psi\left(\frac{1}{2}\right)\right)\nonumber\right.\\&\left.+8\kappa\operatorname{Re}\left(i\psi^{(1)}\left(\frac{1}{2}-\frac{i \kappa}{2}\right)\right)-\kappa^2\operatorname{Re}\left(\psi^{(2)}\left(\frac{1}{2}+\frac{i \kappa}{2}\right)\right)\right].
\end{align}
\end{subequations}
This formula represents the leading order contribution obtained perturbatively in $\delta\ll1$ and $t=(T_c-T)/T_c\ll1$. The profile of the dimensionless function $f(\kappa)$ is shown on Fig. 14. It displays a nonmonotonic dependence with the gradual increase followed by a fall off with the maximum at $\kappa\sim1$. Eq. \eqref{eq:eta} is the main result of this work. For $\kappa\gg1$, or equivalently $\alpha_{R} p_F\gg T_c$, using the proper asymptote of the digamma function $\psi$, one finds  
\begin{equation}
\eta\approx \frac{2}{\sqrt{21\zeta(3)}}
\frac{\delta}{\kappa^2}\left(\frac{\left[\mathbf{h} \times \hat{\mathbf{q}}\right] \cdot \mathbf{c}}{\pi T_c}\right) \sqrt{\frac{T_c-T}{T_c}}.\label{etaLargeSOC}
\end{equation}
The numerical prefactor here takes the value $\approx 0.2815$. In the opposite limit, $\kappa\ll1$, or equivalently $\alpha_{R} p_F\ll T_c$, one finds 
\begin{equation}
\eta\approx \frac{635\zeta(7)}{56\sqrt{42\zeta^3(3)}} \; \delta \kappa^2 
\left(\frac{\left[\mathbf{h} \times \hat{\mathbf{q}}\right] \cdot \mathbf{c}}{\pi T_c}\right) \sqrt{\frac{T_c-T}{T_c}},
\end{equation}
with the numerical value of the prefactor $\approx1.3387$.

\subsection{Strong SOC Limit and the Approximate Inversion Symmetry}

In the limit of strong SOC and large \(p_F\), as considered in \cite{YuanFu22,Yuan23,Yuan23II,IlicBergeret22}, the interband pairing tends to zero and can therefore be neglected. This greatly simplifies the computation as one can work in the helical basis to obtain closed form expressions for the GL coefficients:
\begin{align}
\alpha(\mathbf{q})&=\left[\sum_{\lambda}\int \frac{\nu_{\lambda}}{4\pi}\left[-\ln\frac{1.13\Lambda}{T}+\digamma\left(\frac{\delta\xi_{\lambda}}{T}\right)\right]d\theta_\mathbf{k}\right]^{T_c,\mathbf{q}=0}_T \label{alphaInfSOC}\\
\beta(\mathbf{q})&=\int\sum_{\lambda}\frac{\nu_\lambda}{128\pi^3T^2}\psi^{(2)}\left(\frac{1}{2}-\frac{i\delta\xi_{\lambda}}{4\pi T}\right)d\theta_\mathbf{k}\label{betaInfSOC}
\end{align}
where
\[\digamma\left(x\right)=\text{Re}\left[\psi\left(\frac{1}{2}+\frac{i x}{2\pi}\right)-\psi\left(\frac{1}{2}\right)\right]\,,\]
\(\psi(x)\) is the digamma function, \(\psi^{(2)}(x)=\partial_x^2\psi(x)\), and
\begin{align}
    \delta\xi_{\lambda}(\mathbf{\theta_k;q})&=\xi_{\lambda}(\mathbf{k+q}/2)-\xi_{\lambda'}(-\mathbf{k+q}/2)\nonumber\\
    &=2\lambda h \sin\theta_\mathbf{k}+v_F q \cos(\theta_\mathbf{k}-\theta_\mathbf{q})
\end{align}
The equations, which also neglect corrections to the form of the gap function due to simultaneous presence of the magnetic field and finite momentum pairing (discussed below), are valid to leading order in \(h/(\alpha_R p_F)\), \(q/p_F\), and \(\alpha_R/v_F\). Expanding in \(q\) and carrying out the elementary integrations reproduces the leading terms in \(\alpha_n\) and \(\beta_n\) obtained above.

We now prove that both \(\alpha(\mathbf{q})\) and \(\beta(\mathbf{q})\) as given in Eq. (\ref{alphaInfSOC}-\ref{betaInfSOC}) (and thus also \(F(\mathbf{q})\)) are approximately symmetric under \(\mathcal{I}':q+q_0\rightarrow-q+q_0\) to linear order in \(h\). First, we note that
\begin{align}
    \alpha(\mathbf{q})&=\sum_{\lambda,n} \int \nu_\lambda A_n[2\lambda h \sin\theta_\mathbf{k}+v_F q \cos(\theta_\mathbf{k}-\theta_\mathbf{q})]^{2n}d\theta_\mathbf{k}\nonumber\\
    &\approx\sum_{\lambda,n} \int \nu_\lambda A_n[v_F q \cos(\theta_\mathbf{k}-\theta_\mathbf{q})]^{2n-1}\times\nonumber\\
    &\quad\quad\times[4n\lambda h \sin\theta_\mathbf{k}+v_F q \cos(\theta_\mathbf{k}-\theta_\mathbf{q})]d\theta_\mathbf{k}\,,
\end{align}
where the second line is valid to linear order in \(h\). It follows that, assuming as we later confirm that \(q_0\) is itself of order \(h\),
\begin{align}
\alpha&(-\mathbf{q+2q}_0)\approx\sum_{\lambda,n} \int \nu_\lambda A_n[v_F q \cos(\theta_\mathbf{k}-\theta_\mathbf{q})]^{2n-1}\times\nonumber\\
    &\times[-4n\lambda h \sin\theta_\mathbf{k}+v_F (q+4n q_0) \cos(\theta_\mathbf{k}-\theta_\mathbf{q})]d\theta_\mathbf{k}=\nonumber\\
    &=\sum_{n} \int A_n[v_F q \cos(\theta_\mathbf{k}-\theta_\mathbf{q})]^{2n-1}\times\nonumber\\
    &\times[-4n \Delta \nu h \sin\theta_\mathbf{k}+\nu v_F (q+4n q_0) \cos(\theta_\mathbf{k}-\theta_\mathbf{q})]d\theta_\mathbf{k}\,.
\end{align}
Note furthermore that
\[\int_0^{2\pi}\cos^{2n-1}(\theta-\theta')\sin\theta d\theta=\sin\theta'\int_0^{2\pi}\cos^{2n}(\theta-\theta')d\theta_\mathbf{k}\,.\]
Consequently, \(\alpha(-\mathbf{q+q}_0)=\alpha(\mathbf{q+q}_0)\) to linear order in \(h\) if \(q_0=-(\Delta \nu/\nu) h \sin\theta_\mathbf{q}\). The same proof applies to \(\beta(\mathbf{q})\), and subsequently \(F(-\mathbf{q+q}_0)=F(\mathbf{q+q}_0)+O(h^3)\).

We observe that two features of the Rashba superconductor are vital for the proof to go through. First, it is crucial that the Fermi velocities are equal for the two helical bands at the Fermi momenta. Second, in the case of general SOC, \(H_{\text{SO}}=\mathbf{g}(\mathbf{k})\cdot \boldsymbol{\sigma}\), the term \(\lambda h \sin\theta_\mathbf{k}\) is to be replaced by \(\lambda\mathbf{h}\times\hat{\mathbf{g}}(\mathbf{k})\). The proof of approximate inversion symmetry of the GL free energy only holds if \(\mathbf{h}\times\hat{\mathbf{g}}(\mathbf{k})\propto \cos(\theta_\mathbf{k}-\theta_0)\). It follows that the only type of SOC for which \(\mathcal{I}'\) is an approximate symmetry in the limit of strong SOC and large \(p_F\) is precisely the Rashba-type SOC (including the possibility of a radial Rashba SOC).

Secondly, in the above calculation we also assume the form of the order parameter is the same as that considered in \cite{IlicBergeret22}, which is also key to preserving the approximate inversion symmetry: this is the only pairing that results in equal intraband pairing order parameters on both helical bands. When the intraband pairing is unequal on different bands, both \(\alpha(\mathbf{q})\) and \(\beta(\mathbf{q})\) are approximately symmetric but under two different inversion symmetries about different momenta \(q_0\) and \(q'_0\), such that \(F(\mathbf{q})\) is not approximately symmetric under any inversion symmetry (see Appendix \ref{AppE}). In other words, a great degree of fine-tuning is necessary to obtain the approximate inversion symmetry even in the extreme strong SOC limit. Importantly, the order parameter assumed in \cite{IlicBergeret22} is not in general self-consistent and does not correspond to pure \(s\)-wave spin-singlet pairing in the presence of the in-plane magnetic field and finite momentum pairing. However, although this leads to corrections to the superconducting diode coefficient even if only intraband pairing is assumed, as we discuss in Appendix \ref{AppE} the intraband corrections turn out to be of higher order in \(h/\kappa\), and corrections due to interband pairing turn out to be of lower order. Both types (inter- and intra-band) of such symmetry-breaking corrections are included in our result for \(s\)-wave singlet pairing in Eq. (\ref{eq:eta}) and explain why we find a non-zero superconducting diode effect.


\section{Discussion}\label{SecDisc}

In this work, we have examined the supercurrent diode effect in a paradigmatic model of a superconductor with the Rashba-type spin orbit coupling and Zeeman spin splitting in the clean limit. For this purpose, we utilized both a phenomenological method grounded in the Ginzburg-Landau formalism and a microscopic approach employing diagram techniques. By doing so, we reconcile results of prior works and clarify previously considered limiting cases. Furthermore, we extend our calculations to parameter regimes not previously explored in the literature for this problem.  In particular, our main result defined by Eqs. \eqref{eq:eta}--\eqref{eq:eta-f} gives diode coefficient for an arbitrary relation between the spin-orbit splitting $\Delta_{\text{SO}}=\alpha_R p_F$ and critical temperature $T_c$.  

While our results are perturbative with respect to the strength of Zeeman spin splitting, $\Delta_{\text{Z}}\propto B$, we are able to make rather generic statements about the expected behavior of the supercurrent diode coefficient 
across varying values of spin-orbit splitting $\Delta_{\text{SO}}$ relative to $\Delta_{\text{Z}}$. Specifically, we argue that the supercurrent diode effect is suppressed in both limits: $\Delta_{\text{Z}}/\Delta_{\text{SO}}\ll1$ and $\Delta_{\text{Z}}/\Delta_{\text{SO}}\gg1$. This suppression arises due to the interplay between spin-orbit and Zeeman interactions, influencing the spin texture of spin-split subbands, which is determined by the ratio $\Delta_{\text{Z}}/\Delta_{\text{SO}}$. For $\Delta_{\text{Z}}/\Delta_{\text{SO}}<1$, the spin texture exhibits nontrivial winding, whereas for $\Delta_{\text{Z}}/\Delta_{\text{SO}}>1$, the spin texture is topologically trivial, see Fig. \ref{fig:HelicalBands}. We note that the two-band semiclassical approach assumes that $\Delta_{\text{SO}}$ exceeds the Zeeman splitting, as well as $v_Fq$. The condition $\Delta_{\text{Z}}/\Delta_{\text{SO}}=1$ denotes the point of topological transition where the maximal effect is anticipated. This effect is further magnified by singularities in the density of states resulting from the touching of spin-split Fermi surfaces accompanying this transition. We expect this singularity to be smeared either by the finite-temperature effect or impurity scattering, but enhancement factor to survive. Additionally, our recent work on superconducting multilayers \cite{Osin2024} and ongoing considerations of the Rashba model suggest that SDE survives in the impure superconductor, however in diffusive limit it is further suppressed in $T_c\tau_{\text{el}}\ll1$, where $\tau_{\text{el}}$ is elastic scattering time on impurities.  

 
\section*{Acknowledgments}

This work was financially supported by the National Science Foundation, Quantum Leap Challenge Institute for Hybrid Quantum Architectures and Networks Grant No. OMA-2016136. A. L. gratefully acknowledges H. I. Romnes Faculty Fellowship provided by the University of Wisconsin-Madison Office of the Vice Chancellor for Research and Graduate Education with funding from the Wisconsin Alumni Research Foundation. M. K. acknowledges financial support from the Israel Science Foundation, Grant No. 2665/20. 

\appendix

\section{Auxiliary equalities}\label{AppA}

We provide a list of equalities in this Appendix that will be applied to the subsequent evaluation of the diagrams. The following are some angle integrals:
\begin{equation}\label{eq:AppA5}
\begin{aligned}
S_{m j}^{(\mu \nu)} &=\int \frac{d \hat{p}}{2 \pi} \operatorname{Tr}\left\{\boldsymbol{\sigma}_m \Pi^{(\mu)}(\mathbf{p}) v_j(\mathbf{p}) \Pi^{(\nu)}(\mathbf{p})\right\} \\
& =\frac{1}{2} \varepsilon_{m j s} c^s\left(\begin{array}{cc}
\frac{p}{m}+\alpha_R & \alpha_R \\
\alpha_R & -\frac{p}{m} +\alpha_R
\end{array}\right)^{(\mu \nu)},
\end{aligned}
\end{equation}
\begin{equation}\label{eq:AppA6}
\begin{aligned}
P_{i j}^{(\mu \nu)}= & \int \frac{d \hat{p}}{2 \pi} \operatorname{Tr}\left\{v^i(\mathbf{p}) \Pi^{(\mu)}(\mathbf{p}) v^j(\mathbf{p}) \Pi^{(\nu)}(\mathbf{p})\right\} \\
= & \frac{1}{2}\delta^{\perp}_{i j} \left(\begin{array}{cc}
\left(\frac{p}{m} +\alpha_R\right)^2 & \alpha^2_R \\
\alpha^2_R & \left(\frac{p}{m} -\alpha_R\right)^2
\end{array}\right)^{(\mu \nu)},
\end{aligned}
\end{equation}
\begin{widetext}
\begin{equation}\label{eq:AppA7}
\begin{aligned}
& P_{i j k l}^{(\mu \nu \beta \lambda)}=\int \frac{d \hat{p}}{2 \pi} \operatorname{Tr}\left\{v^i(\mathbf{p}) \Pi^{(\mu)}(\mathbf{p}) v^j(\mathbf{p}) \Pi^{(\nu)}(\mathbf{p}) v^k(\mathbf{p}) \Pi^{(\beta)}(\mathbf{p}) v^l(\mathbf{p}) \Pi^{(\lambda)}(\mathbf{p})\right\} \\
& = \frac{1}{64}\left(\delta_{i j}^\perp \delta_{k l}^\perp+\delta_{i k}^\perp \delta_{j l}^\perp+\delta_{i l}^\perp \delta_{j k}^\perp\right)\left(\frac{p}{m}\right)^3\left[\left(\frac{p}{m}\right)(1+\mu \nu+\mu \beta+\right.+\mu \lambda+\nu \beta+\nu \lambda+\beta \lambda+\mu \nu \beta \lambda) +4 \alpha_R(\mu+\nu+\beta+\\&\left.+\lambda+\mu \nu \beta+\mu \nu \lambda+\mu \beta \lambda+\nu \beta \lambda)\right]+O\left(\alpha^2_R\right),
\end{aligned}
\end{equation}
where we have introduced the notation, $\delta^{\perp}_{i j}=\delta_{i j}-c_i c_j$.
\begin{equation}\label{eq:AppA8}
\begin{aligned}
&S_{mijk}^{(\mu \nu \beta \lambda)} =\int \frac{d \hat{p}}{2 \pi} \operatorname{Tr}\left\{\boldsymbol{\sigma}_m \Pi^{(\mu)}(\mathbf{p}) v_i(\mathbf{p}) \Pi^{(\nu)}(\mathbf{p})v_j(\mathbf{p}) \Pi^{(\beta)}v_k(\mathbf{p}) \Pi^{(\lambda)}(\mathbf{p})\right\}
\\&= \frac{1}{64}\left(\frac{p}{m}\right)^3(\mu+\nu+\beta+\lambda+\mu \nu \beta+\mu \nu \lambda+\mu \beta \lambda+\nu \beta \lambda) c^a\left(\varepsilon_{a m k} \delta_{i j}^{\perp}+\varepsilon_{a m i} \delta_{j k}^{\perp}+\varepsilon_{a m j} \delta_{i k}^{\perp}\right) \\
& +\frac{1}{16} \alpha_R\left(\frac{p}{m}\right)^2 c^l \biggl[ \left(\varepsilon_{l m k} \delta_{i j}^{\perp}+\varepsilon_{l m i} \delta_{j k}^{\perp}+\varepsilon_{l m j} \delta_{i k}^{\perp}\right)+\mu \nu\left(\frac{3}{2} \varepsilon_{l m k} \delta_{i j}^{\perp}-\frac{1}{2} \varepsilon_{l m i} \delta_{j k}^{\perp}+\frac{3}{2} \varepsilon_{l m j} \delta_{i k}^{\perp}\right)+2\mu \beta \varepsilon_{l m k} \delta_{i j}^{\perp}\\
& +\mu \lambda\left(\frac{1}{2} \varepsilon_{l m k} \delta_{i j}^{\perp}+\frac{1}{2} \epsilon_{l m i} \delta_{j k}^{\perp}+\frac{1}{2} \varepsilon_{l m j} \delta_{i k}^{\perp}\right) \left.+\nu \beta\left(\frac{3}{2} \varepsilon_{l m k} \delta_{i j}^{\perp}+\frac{3}{2} \varepsilon_{l m i} \delta_{j k}^{\perp}-\frac{1}{2} \varepsilon_{l m j} \delta_{i k}^{\perp}\right)\right.\\
&+2\nu \lambda \varepsilon_{l m i} \delta_{j k}^{\perp}+\beta \lambda\left(-\frac{1}{2} \varepsilon_{l m k} \delta_{i j}^{\perp}+\frac{3}{2} \varepsilon_{l m i} \delta_{j k}^{\perp}+\frac{3}{2} \varepsilon_{l m j} \delta_{i k}^{\perp}\right)+\mu \nu \beta \lambda \left(2 \varepsilon_{l m j} \delta_{i k}^{\perp}\right)\biggr] +\order{\alpha^2_R},
\end{aligned}
\end{equation}
\end{widetext}
The radial integrals below should be evaluated up to terms linear in $\delta=\alpha_R/v_F$. Because energy bands of electrons with positive and negative helicities are different, $\xi_{( \lambda)}(p)=\xi +\lambda \alpha_R p \; (\xi=\frac{p^2}{2 m}-\mu)$ with $\lambda=\pm 1$, there are two Fermi momenta at a given $p$, whose values with the adopted accuracy are
$p_{\lambda}\approx p_F[1 -\lambda \delta]$. 
Near these momenta, the energy branches behave as
$\xi_{( \lambda)}(p) \cong \xi[1 + \lambda \delta] +\lambda \alpha_R p_F$.
Also, with the same accuracy,
$\frac{p(\xi)}{p_F} \cong1-\lambda \delta+\frac{\xi_{(\lambda)}}{v_F p_F}$ for $\left|p-p_{\lambda}\right| \ll p_F$, 
and $\frac{d \xi}{d \xi_{( \lambda)}} \cong[1 -\lambda \delta]$.
The above relations allow one to show the validity of the following integrals up to $\order{\delta}$:
\begin{subequations}
\begin{equation}
\int \frac{d \xi}{2 \pi} G_{(\mu)}^2 \left[G_{(\mu)}^{\mathrm{(r)}}\right]^2= 2\frac{1- \mu \delta}{\left(2\left|\omega_n\right|\right)^3},
\end{equation}
\begin{equation}
\int \frac{d \xi}{2 \pi} G_{(\mu)}^3 G_{(\mu)}^{\mathrm{(r)}}\left(\frac{p}{m}\right)^2= v_F^2\frac{1- 3\mu \delta}{\left(2\left|\omega_n\right|\right)^3},
\end{equation}
\begin{equation}
\int \frac{d \xi}{2 \pi} G_{(\mu)}^5 G_{(\mu)}^{\mathrm{(r)}}\left(\frac{p}{m}\right)^4= v_F^4\frac{1- 5\mu \delta}{\left(2\left|\omega_n\right|\right)^5},
\end{equation}
\begin{equation}
\int \frac{d \xi}{2 \pi}\left[G_{(\mu)}^{(\text {r})}\right]^4 G_{(\mu)}^2\left(\frac{p}{m}\right)^2= -4 v_F^2 \frac{1-3 \mu \delta}{\left(2\left|\omega_n\right|\right)^5}
\end{equation}
\end{subequations}
Similarly, the following integrals are valid up to $\order{\delta^0}$:
\begin{equation}
\int \frac{d \xi}{2 \pi} G_{(\mu)} G_{(\mu)}^{\mathrm{(r)}}G_{(-\mu)} G_{(-\mu)}^{\mathrm{(r)}}= \frac{1}{4\left|\omega_n\right|[\left|\omega_n\right|^2+(\alpha_R p_F)^2]}.
\end{equation}


\section{Anomalous diagrams in Figs. \ref{fig7}, \ref{fig8} and \ref{fig10}}

In this Appendix, we evaluate anomalous contributions to the thermodynamic potential $\Phi_\mathrm{2.an}$ and $\Phi_\mathrm{4.an}$ represented by the diagrams in Figs. \ref{fig7}, \ref{fig8} and \ref{fig10}. Each of these diagrams can be decomposed into a product of two factors. The first of these $S$ (it will be called the slow factor) stems from the integration of functions that slowly vary in coordinate space: the magnetic field $\mathbf{B}(\mathbf{q}^{\prime})$, and the gap function $\Delta(\mathbf{q})$. Arguments of these functions in momentum space will be called slow variables. The second factor $Q$ (it will be called the quick factor) appears as a result of integration over the momentum $\mathbf{p}$ (or $\mathbf{p}$ and $\mathbf{p}^{\prime}$) of the fermion lines, taking a trace (in the clockwise direction) over spin variables of all entities entering an electron loop and summing over the fermion frequency $\omega_n$. 

\subsection{Evaluation of the diagram in Figs. \ref{fig7}$(1q)$ and \ref{fig8}$(1q)$}

The slow factor $S_{\mathrm{2.an(1q)}}$ common in Figs. \ref{fig7}$(1q)$ and \ref{fig8}$(1q)$ is given by
\begin{equation}
S_{\mathrm{2.an(1q)}}=\int_{\mathbf{q}} h_m |\Delta\left(\mathbf{q}\right)|^2 q_j,
\end{equation}
where $\mathbf{h}=\mu_B \mathbf{B}$. Its quick factor $Q_{\mathrm{2.an(1q)}}$ is given by
\begin{equation}
\begin{aligned}
Q_{\mathrm{2.an(1q)}}= & T \sum_{\omega_n} \int_{\mathbf{p}} \operatorname{Tr}\{\boldsymbol{\sigma}_m G\left(i \omega_n, \mathbf{p}\right)
G^{(\mathrm{r})}\left(i \omega_n, \mathbf{p}\right) \\&\times v_j(\mathbf{p}) G^{(\mathrm{r})}\left(i \omega_n, \mathbf{p}\right) G\left(i \omega_n, \mathbf{p}\right)\}
\end{aligned}
\end{equation}
where
\begin{equation}
\int_{\mathbf{p}}=2\pi\nu \int \frac{d \xi}{2 \pi} \int \frac{d \hat{p}}{2 \pi}.
\end{equation}
The main component of $Q_{\mathrm{2.an(1q)}}$ is the function,
\begin{equation}
\begin{aligned}
I_{\mathrm{2.an(1q)}}&\left(\omega_n\right)=2\pi\nu \int \frac{d \xi}{2 \pi} \int \frac{d \hat{p}}{2 \pi} \operatorname{Tr}\{\boldsymbol{\sigma}_m G\left(i \omega_n, \mathbf{p}\right) \\&\times G^{(\mathrm{r})}\left(i \omega_n, \mathbf{p}\right) v_j(\mathbf{p}) G^{(\mathrm{r})}\left(i \omega_n, \mathbf{p}\right) G\left(i \omega_n, \mathbf{p}\right)\}\\
=&2\pi\nu\int \frac{d \xi}{2 \pi} \sum_{\mu \nu} S_{m j}^{(\mu \nu)}R_{(\mu \nu)},
\end{aligned}
\end{equation}
where $S_{m j}^{(\mu \nu)}$ is defined by Eq. (\ref{eq:AppA5}) and $R_{(\mu \nu)}$ is given by
\begin{equation}
R_{(\mu \nu)}=G_{(\mu)} G_{(\mu)}^{\text {(r)}} G_{(\nu)} G_{(\nu)}^{(\text{r})}.
\end{equation}
Elements of $S_{mj}^{(\mu \nu)} R_{(\mu \nu)}$ diagonal in indices $\mu$ and $\nu$ contribute to $I_{\mathrm{2.an(1q)}}(\omega_n):$
\begin{equation}
(2\pi\nu)\frac{1}{2} \varepsilon_{m j i} c^i \frac{(-4 \alpha_R)}{\left(2\left|\omega_n\right|\right)^3},
\end{equation}
while the off-diagonal elements contribute
\begin{equation}
(2\pi\nu)\frac{1}{2} \varepsilon_{m j i} c^i \frac{1}{2\left|\omega_n\right|} \cdot \frac{4 \alpha_R}{(2|\omega_n|)^2+(2\alpha_R p_F)^2}.
\end{equation}
Combining these two contributions, we obtain for $I_{\mathrm{2.an(1q)}}\left(\omega_n\right)$:
\begin{equation}
I_{\mathrm{2.an(1q)}}\left(\omega_n\right)= \frac{4\pi\nu \alpha_R \varepsilon_{j m i} c^i\left(2 \alpha p_F\right)^2}{(2|\omega_n|)^3[(2|\omega_n|)^2+(2 \alpha_R p_F)^2]},
\end{equation}
from which we get for the free energy:
\begin{equation}
\Delta\Omega=\frac{1}{2}\Phi_{\mathrm{2.an(1q)}}=S_{\mathrm{2.an(1q)}}Q_{\mathrm{2.an(1q)}}=\int_{\mathbf{q}}\alpha_1 q |\Delta\left(\mathbf{q}\right)|^2,
\end{equation}
where the contribution from two diagrams in Figs. \ref{fig7}$(1q)$ and \ref{fig8}$(1q)$ being equal in magnitude cancels the factor of $\frac{1}{2}$ and thus we get for $\alpha_1$:
\begin{equation}
\alpha_1=-\frac{\nu\alpha_R}{2}(\left[\mathbf{h} \times \hat{\mathbf{q}}\right] \cdot \mathbf{c})\;  \sum_{\omega_n>0} \frac{2\pi T\left(\alpha_R p_F\right)^2}{ \omega_n^3[ \omega_n^2+\left(\alpha_R p_F\right)^2]}.
\end{equation}

\subsection{Evaluation of the diagram in Fig. \ref{fig7}$(3q)$ and \ref{fig8}$(3q)$}

The slow factor $S_{\mathrm{2.an(3q)}}$ common in Figs. \ref{fig7}$(3q)$ and \ref{fig8}$(3q)$ is given by
\begin{equation}
S_{\mathrm{2.an(3q)}}=\int_{\mathbf{q}} h_m |\Delta\left(\mathbf{q}\right)|^2 q_i q_j q_k,
\end{equation}
Its quick factor $Q_{\mathrm{2.an(3q)}}$ is given by
\begin{equation}
\begin{aligned}
& Q_{\mathrm{2.an(3q)}}= T \sum_{\omega_n} \int_{\mathbf{p}}
\operatorname{Tr}\left[G \boldsymbol{\sigma}^m G G^{\mathrm{(r)}}  v^i(\mathbf{p}) \right. \\
&\left.\times G^{\mathrm{(r)}} v^j(\mathbf{p}) G^{\mathrm{(r)}} v^k(\mathbf{p}) G^{\mathrm{(r)}} \right],
\end{aligned}
\end{equation}
where all the $G$'s and $G^{\mathrm{(r)}}$'s have the same argument $\left(i \omega_n, \mathbf{p}\right)$, suppressed for brevity.
The main component of both diagrams is the function,
\begin{equation}
\begin{aligned}
I_{\mathrm{2.an(3q)}}\left(\omega_n\right)&=  h_m q_i q_j q_k \int 2\pi\nu \frac{d \xi}{2 \pi} \int \frac{d \hat{p}}{2 \pi} \operatorname{Tr}\left[G \boldsymbol{\sigma}^m G  \right. \\
&\left.\times G^{\mathrm{(r)}} v^i(\mathbf{p}) G^{\mathrm{(r)}} v^j(\mathbf{p}) G^{\mathrm{(r)}} v^k(\mathbf{p}) G^{\mathrm{(r)}} \right]\\
&=h_m q_i q_j q_k\int 2\pi\nu \frac{d \xi}{2 \pi}\sum_{\mu \nu \beta \lambda} S_{mijk}^{(\mu \nu \beta \lambda)}L_{(\mu \nu \beta \lambda)},
\end{aligned}
\end{equation}
where $S_{mijk}^{(\mu \nu \beta \lambda)}$ is defined by Eq. (\ref{eq:AppA8}) and $L_{(\mu \nu \beta \lambda)}$ is given by
\begin{equation}
L_{(\mu \nu \beta \lambda)}=G_{(\mu)} G_{(\mu)}^{\mathrm {(r)}} G_{(\nu)}^{(\mathrm{r})} G_{(\beta)}^{(\mathrm{r})} G_{(\lambda)} G_{(\lambda)}^{\mathrm {(r)}}.
\end{equation}
Here, elements of $h_m q_i q_j q_k S_{mijk}^{(\mu \nu \beta \lambda)}L_{(\mu \nu \beta \lambda)}$ for different values of the indices $\mu$, $\nu$, $\beta$ and $\lambda$ contribute to $I_{\mathrm{2.an(3q)}}(\omega_n)$ in the following way:
\begin{equation}
\begin{aligned}
&I_{\mathrm{2.an(3q)}}(\omega_n)=2\pi\nu\alpha_R v_F^2 q^2 (\left[\mathbf{h} \times \mathbf{q}\right] \cdot \mathbf{c}) \\
&\times \begin{cases}
\frac{3}{32\left|\omega_n\right|^5}, & \mu=\nu=\beta=\lambda\\
-\frac{1}{32\left|\omega_n\right|\left[\left|\omega_n\right|^2+(\alpha_R p_F)^2\right]^2}, &\mu=\nu\neq \beta=\lambda\\
-\frac{(\alpha_R p_F)^4+(\alpha_R p_F)^2|\omega_n|^2+4|\omega_n|^4}{64\left|\omega_n\right|^3\left[\left|\omega_n\right|^2+(\alpha_R p_F)^2\right]^3}, &\overset{\mu=\nu=\beta\neq\lambda}{\mu\neq\nu=\beta=\lambda}
\end{cases}
\end{aligned}
\end{equation}
Thus we obtain for $I_{\mathrm{2.an(3q)}}(\omega_n)$ combining all three contributions:
\begin{align}
&I_{\mathrm{2.an(3q)}}(\omega_n)=2\pi\nu\alpha_R v_F^2 q^2 (\left[\mathbf{h} \times \mathbf{q}\right] \cdot \mathbf{c})\times \nonumber \\ 
&\frac{15\left|\omega_n\right|^4 (\alpha_R p_F)^2+17\left|\omega_n\right|^2 (\alpha_R p_F)^4+6 (\alpha_R p_F)^6}{64\left|\omega_n\right|^5\left[\left|\omega_n\right|^2+(\alpha_R p_F)^2\right]^3},
\end{align}
from which we get for the corresponding term in the free energy:
\begin{equation}
\Delta\Omega=S_{\mathrm{2.an(3q)}}Q_{\mathrm{2.an(3q)}}=\int_{\mathbf{q}}\alpha_3 q^3 |\Delta\left(\mathbf{q}\right)|^2,
\end{equation}
where the contribution from two diagrams in Figs. \ref{fig7}$(3q)$ and \ref{fig8}$(3q)$ being equal in magnitude cancels the factor of $\frac{1}{2}$ and thus we get for $\alpha_3$:
\begin{equation}
\begin{split}
&\alpha_3=\frac{\alpha_R \nu v_F^2 }{32}(\left[\mathbf{h} \times \hat{\mathbf{q}}\right] \cdot \mathbf{c}) \; \\&\times2\pi T \sum_{\omega_n>0} \frac{15 \omega_n^4 \left(\alpha_R p_F\right)^2+17\omega_n^2 \left(\alpha_R p_F\right)^4+6\left(\alpha_R p_F\right)^6}{ \omega_n^5\left[ \omega_n^2+\left(\alpha_R p_F\right)^2\right]^3}.
\end{split}
\end{equation}

\subsection{Evaluation of the diagram in Figs. \ref{fig10}$(1q)a$ and \ref{fig10}$(1q)b$}

The slow factor $S_{\mathrm{4.an(1q)}}$ in the diagram in Fig. \ref{fig10}$(1q)a$ is given by
\begin{equation}
S_{\mathrm{4.an(1q)}}=\int_{\mathbf{q}} h_m |\Delta\left(\mathbf{q}\right)|^4 q_j,
\end{equation}
Its quick factor $Q_{\mathrm{4.an(1q)}}$ is given by
\begin{equation}
\begin{aligned}
Q_{\mathrm{4.an(1q)}}= & T \sum_{\omega_n} \int_{\mathbf{p}} \operatorname{Tr}\left[G \boldsymbol{\sigma}^m G G^{\mathrm{(r)}}  v^j(\mathbf{p})G^{\mathrm{(r)}}  G G^{\mathrm{(r)}}\right]\\
=2\pi\nu \int\frac{d \xi}{2 \pi}&\frac{d \hat{p}}{2 \pi} \sum_{\mu,\nu=\pm} \operatorname{Tr}\left[\boldsymbol{\sigma}_m \Pi^{(\mu)}(\mathbf{p}) v_j(\mathbf{p}) \Pi^{(\nu)}(\mathbf{p})\right]\\
&\times G_{(\mu)} G_{(\mu)}^{(\mathrm{r})} G_{(\nu)}^2\left[G_{(\nu)}^{(\mathrm {r})}\right]^2\\
=&2\pi\nu\int\frac{d \xi}{2 \pi} \sum_{\mu,\nu=\pm} S_{m j}^{(\mu \nu)}G_{(\mu)} G_{(\mu)}^{(\mathrm{r})} G_{(\nu)}^2\left[G_{(\nu)}^{(\mathrm {r})}\right]^2,
\end{aligned}
\end{equation}
where $S_{m j}^{(\mu \nu)}$ is defined by Eq. (\ref{eq:AppA5}),
from which we get for the free energy:
\begin{equation}
\Delta\Omega=2S_{\mathrm{4.an(1q)}}Q_{\mathrm{4.an(1q)}}=\int_{\mathbf{q}}\beta_1 q |\Delta\left(\mathbf{q}\right)|^4,
\end{equation}
where the contribution from two diagrams in Figs. \ref{fig10}$(1q)a$ and \ref{fig10}$(1q)b$ are equal in magnitude and there are four such group of diagrams, 8 diagrams in total and thus we get for $\beta_1$:
\begin{align}
&\beta_1=\frac{\nu\alpha_R}{4}(\left[\mathbf{h} \times \hat{\mathbf{q}}\right] \cdot \mathbf{c})\nonumber \\ 
&\times 2\pi T \sum_{\omega_n>0} \frac{3\left(\alpha_R p_F\right)^4+5\left(\alpha_R p_F\right)^2\omega_n^2}{ \omega_n^5\left[\omega_n^2+\left(\alpha_R p_F\right)^2\right]^2},
\end{align}


\section{Conventional diagrams in Figs. \ref{fig6} and \ref{fig9}}

In this Appendix, we evaluate $\Phi_\mathrm{2.c}$ and $\Phi_\mathrm{4.c}$ represented by the conventional diagrams in Figs. \ref{fig6} and \ref{fig9} respectively. 

\subsection{Evaluation of the diagram in Fig. \ref{fig6}$(0q)$}
The slow factor $S_{\mathrm{2.c(0q)}}$ in the diagram in Fig. \ref{fig6}$(0q)$ is given by
\begin{equation}
S_{\mathrm{2.c(0q)}}=\int_{\mathbf{q}} |\Delta\left(\mathbf{q}\right)|^2,
\end{equation}
Its quick factor $Q_{\mathrm{2.c(0q)}}$ is given by
\begin{equation}
\begin{aligned}
Q_{\mathrm{2.c(0q)}}= & T \sum_{\omega_n} \int_{\mathbf{p}} \operatorname{Tr}[G\left(i \omega_n, \mathbf{p}\right)
G^{(\mathrm{r})}\left(i \omega_n, \mathbf{p}\right)]\\
=&\int \nu d \xi \int \frac{d \hat{p}}{2 \pi}\sum_\mu T \sum_{\omega_n} G_{(\mu)} G_{(\mu)}^{(\mathrm{r})}
\end{aligned}
\end{equation}
Changing the order of summation and integration we get
\begin{equation}
\begin{aligned}
Q_{\mathrm{2.c(0q)}}& =-\sum_{\mu= \pm 1} \nu[1-\mu \delta] \int_0^{\omega_D} \frac{d \xi}{\xi} \tanh \left(\frac{\xi}{2T}\right)\\&=-2\nu\ln \frac{T_c}{T}-\frac{2}{\lambda_s}+\order{\delta^2},
\end{aligned}
\end{equation}
where $\omega_D$ is the Debye frequency. 
We have used as usual \cite{Abrikosov}, $\frac{\Delta_0 \gamma_E}{\pi}=T_c$ and $\frac{1}{\lambda_s}=\nu \ln \frac{2 \omega_D}{\Delta_0}$, with $\ln\gamma_E=0.5772 \dots$ as the Euler's constant. In a proximity of the transition, where $T_c-T \ll T_c$, the diagram in Fig. \ref{fig6}$(0q)$ yields
\begin{equation}
\Phi_{\mathrm{2.c(0q)}}=-\int_{\mathbf{q}} |\Delta\left(\mathbf{q}\right)|^2\left[2\nu\left(\frac{T_c-T}{T_c}\right)+\frac{2}{\lambda_s}\right],
\end{equation}
This term together with the first term in Eq. (\ref{eq:IIIA8}) yields
\begin{equation}
\frac{1}{\lambda_s}\int_{\mathbf{q}} |\Delta\left(\mathbf{q}\right)|^2+\frac{1}{2}\Phi_{\mathrm{2.c(0q)}}=\int_{\mathbf{q}}\alpha_0 |\Delta\left(\mathbf{q}\right)|^2,
\end{equation}
where
\begin{equation}
\alpha_0=-\nu\left(\frac{T_c-T}{T_c}\right).
\end{equation}

\subsection{Evaluation of the diagram in Figs. \ref{fig6}$(2q)$}

The slow factor $S_{\mathrm{2.c(2q)}}$ in the diagram in Fig. \ref{fig6}$(2q)$ is given by
\begin{equation}
S_{\mathrm{2.c(2q)}}=\int_{\mathbf{q}} |\Delta\left(\mathbf{q}\right)|^2 q_i q_j,
\end{equation}
Its quick factor $Q_{\mathrm{2.c(2q)}}$ is given by
\begin{equation}
\begin{aligned}
Q_{\mathrm{2.c(2q)}}= & T \sum_{\omega_n} \int_{\mathbf{p}} \operatorname{Tr}[Gv_i(\mathbf{p})Gv_j(\mathbf{p})GG^{(\mathrm{r})}]\\
&=T \sum_{\omega_n} \int \nu d \xi \sum_{\mu, \nu= \pm 1} P_{i j}^{(\mu \nu)} G_{(\mu)} G_{(\nu)}^2 G_{(\nu)}^{\mathrm{(r)}}
\end{aligned}
\end{equation}
where $P_{i j}^{(\mu \nu)}$ is defined by Eq. (\ref{eq:AppA6}) and the common argument $\left(i \omega_n, \mathbf{p}\right)$ of all the Green's function under trace was suppressed for compactness. As we can see, only $\mu=\nu$ terms will dominate if we neglect corrections of the order $\order{\delta^2}$. So, only one integral left to do is the $\xi$ integral, $\int d \xi G_{(\mu)}^3 G_{(\mu)}^{\mathrm{(r)}}$ which is tabulated in Appendix A, from which we get for the free energy:
\begin{equation}
\begin{aligned}
\Delta\Omega&=\frac{1}{2}\Phi_{\mathrm{2.c(2q)}}\left(q\right)\\&=S_{\mathrm{2.c(2q)}}Q_{\mathrm{2.c(2q)}}=\int_{\mathbf{q}}\alpha_2 q^2 |\Delta\left(\mathbf{q}\right)|^2,
\end{aligned}
\end{equation}
where
\begin{equation}
\alpha_2=\frac{\nu v_F^2}{8}\sum_{\omega_n>0} \frac{2\pi T }{\omega_n^3}.
\end{equation}

\subsection{Evaluation of the diagram in Figs. \ref{fig6}$(4q)$}

The slow factor $S_{\mathrm{2.c(4q)}}$ in the diagram in Fig. \ref{fig6}$(4q)$ is given by
\begin{equation}
S_{\mathrm{2.c(4q)}}=\int_{\mathbf{q}} |\Delta\left(\mathbf{q}\right)|^2 q_i q_j q_k q_l,
\end{equation}
Its quick factor $Q_{\mathrm{2.c(4q)}}$ is given by
\begin{equation}
\begin{aligned}
&Q_{\mathrm{2.c(4q)}}=\nonumber \\  
&T \sum_{\omega_n} \int_{\mathbf{p}} \operatorname{Tr}[G v_i(\mathbf{p})G v_j(\mathbf{p})G v_k(\mathbf{p})G v_l(\mathbf{p})G G^{(\mathrm{r})}]\\
&=T \sum_{\omega_n} \int  \nu d \xi \sum_{\mu, \nu, \beta, \lambda= \pm 1} P_{i j k l}^{(\mu \nu \beta \lambda)} \\&\times G_{(\mu)}G_{(\nu)}G_{(\beta)} G_{(\lambda)}^2 G_{(\lambda)}^{\mathrm{(r)}}
\end{aligned}
\end{equation}
where all the $G$'s and $G^{\mathrm{(r)}}$'s have the same argument $\left(i \omega_n, \mathbf{p}\right)$ and $P_{i j k l}^{(\mu \nu \beta \lambda)}$ is defined by Eq. (\ref{eq:AppA7}). As we can see, only $\mu=\nu=\beta=\lambda$ terms will dominate if we neglect corrections of the order $\order{\delta^2}$. So, only one integral left to do is the $\xi$ integral, $\int d \xi G_{(\mu)}^5 G_{(\mu)}^{\mathrm{(r)}}$ which is tabulated in Appendix A, from which we get for the free energy:
\begin{equation}
\Delta\Omega=\frac{1}{2}\Phi_{\mathrm{2.c(4q)}}\left(q\right)=S_{\mathrm{2.c(4q)}}Q_{\mathrm{2.c(4q)}}=\int_{\mathbf{q}}\alpha_4 q^4 |\Delta\left(\mathbf{q}\right)|^2,
\end{equation}
where
\begin{equation}
\alpha_4=-\frac{3\nu v_F^4}{128} \sum_{\omega_n>0} \frac{2\pi T}{ \omega_n^5}.
\end{equation}

\subsection{Evaluation of the diagram in Figs. \ref{fig9}$(0q)$}

The slow factor $S_{\mathrm{4.c(0q)}}$ in the diagram in Fig. \ref{fig9}$(0q)$ is given by
\begin{equation}
S_{\mathrm{4.c(0q)}}=\int_{\mathbf{q}} |\Delta\left(\mathbf{q}\right)|^4,
\end{equation}
Its quick factor $Q_{\mathrm{4.c(0q)}}$ is given by
\begin{equation}
\begin{aligned}
&Q_{\mathrm{4.c(0q)}}=T \sum_{\omega_n} \int_{\mathbf{p}} \operatorname{Tr}\left[GG^{(\mathrm{r})}
GG^{(\mathrm{r})}\right]\\
&=T \sum_{\omega_n} \int \nu d \xi \sum_{\mu= \pm 1} G_{(\mu)}^2 \left[G_{(\mu)}^{\mathrm{(r)}}\right]^2\\
&=T \sum_{\omega_n} \sum_{\mu= \pm 1} 2 m \frac{1-\mu \delta}{\left(2\left|\omega_n\right|\right)^3},
\end{aligned}
\end{equation}
from which we get for the free energy:
\begin{equation}
\Delta\Omega=S_{\mathrm{4.c(0q)}}Q_{\mathrm{4.c(0q)}}=\int_{\mathbf{q}}\beta_0 |\Delta\left(\mathbf{q}\right)|^4,
\end{equation}
with the coefficient 
\begin{equation}
\beta_0=\frac{\nu}{4} \sum_{\omega_n>0} \frac{2\pi T}{ \omega_n^3}.
\end{equation}

\subsection{Evaluation of the diagram in Figs. \ref{fig9}$(2q)$}

The slow factor $S_{\mathrm{4.c(2q)}}$ in the diagram in Fig. \ref{fig9}$(2q)a$ is given by
\begin{equation}
S_{\mathrm{4.c(2q)}}=\int_{\mathbf{q}} |\Delta\left(\mathbf{q}\right)|^4 q_i q_j,
\end{equation}
Their quick factor $Q_{\mathrm{4.c(2q)}}$ is given by
\begin{equation}
\begin{aligned}
&Q_{\mathrm{4.c(2q)}}=T \sum_{\omega_n} \int_{\mathbf{p}} \operatorname{Tr}[G G^{(\mathrm{r})}v_i(\mathbf{p})G^{(\mathrm{r})} v_j(\mathbf{p})G^{(\mathrm{r})}G   G^{(\mathrm{r})}]\\
&=T \sum_{\omega_n} \int \nu d \xi \sum_{\mu, \nu= \pm 1} P_{i j}^{(\mu \nu)} G_{(\mu)}^{(\mathrm{r})}G_{(\nu)}^2 \left[G_{(\nu)}^{(\mathrm{r})}\right]^3\\
&=T \sum_{\omega_n} \sum_{\mu= \pm 1}-4\pi\nu v_F^2 \frac{1-3 \mu \delta}{\left(2\left|\omega_n\right|\right)^5}\delta^\perp_{ij}
\end{aligned}
\end{equation}
where all the $G$'s and $G^{\mathrm{(r)}}$'s have the same argument $\left(i \omega_n, \mathbf{p}\right)$ and $P_{i j}^{(\mu \nu)}$ is defined by Eq. (\ref{eq:AppA6}). As we can see, only $\mu=\nu$ terms will dominate (we neglect corrections of the order $\order{\delta^2}$). So, only one integral left to do is the $\xi$ integral, $\int \frac{d \xi}{2 \pi} G_{(\mu)}^2 \left[G_{(\mu)}^{\mathrm{(r)}}\right]^4$, which is tabulated in Appendix A. Notice that there are three diagrams in Figs. \ref{fig9}$(2q)a$, \ref{fig9}$(2q)b$ and \ref{fig9}$(2q)c$, all of which give the same contribution after explicit evaluation, from which we get for the free energy:
\begin{equation}
\Delta\Omega=\frac{3}{4}S_{\mathrm{4.c(2q)}}Q_{\mathrm{4.c(2q)}}=\int_{\mathbf{q}}\beta_2 q^2 |\Delta\left(\mathbf{q}\right)|^4,
\end{equation}
where the coefficient is given by 
\begin{equation}
\beta_2=-\frac{3\nu v_F^2}{16}  \sum_{\omega_n>0} \frac{2\pi T}{ \omega_n^5}. 
\end{equation}


\section{GL coefficients in terms of digamma functions}

Completing the Matsubara summation and writing the coefficients of $\alpha(\mathbf{q})$ and $\beta(\mathbf{q})$ in terms of the digamma function $\psi(x)$ and its $N^{\text{th}}$ order derivatives, 
\begin{equation}
\psi^{(N)}(z)=(-1)^{N+1}N!\sum^{\infty}_{n=0}\frac{1}{(n+z)^{N+1}},
\end{equation}
as well as Riemann zeta-function,  
\begin{equation}
\zeta(s)=\sum^{\infty}_{n=1}\frac{1}{n^s},
\end{equation}
we get for conventional coefficients:
\begin{widetext}
\begin{subequations}
\begin{align}
\alpha_0=-\nu\frac{T_c-T}{T_c},\qquad
\alpha_2=\frac{7\zeta(3) \nu v_F^2}{32 \left(\pi T\right)^2},\qquad
\alpha_4=-\frac{93\zeta\left(5\right)\nu v_F^4}{2048\left(\pi T\right)^4},\\
\beta_0=\frac{7\zeta(3) \nu}{16 \left(\pi T\right)^2},\qquad
\beta_2=-\frac{93\zeta(5) \nu v_F^2}{256 \left(\pi T\right)^4},\qquad
\beta_4=\frac{5715\zeta(7) \nu v_F^4}{32768 \left(\pi T\right)^6},
\end{align}
\end{subequations}
For anomalous coefficients we get,
\begin{subequations}
\begin{align}
&\alpha_1=-\frac{\nu\alpha_R (\left[\mathbf{h} \times \hat{\mathbf{q}}\right] \cdot \mathbf{c})}{\left(\pi T\right)^2} \biggl[\frac{7\zeta\left(3\right)}{8}-\frac{1}{2\kappa^2}\operatorname{Re}\left(\psi\left(\frac{1}{2}+\frac{i \kappa}{2}\right)-\psi\left(\frac{1}{2}\right)\right)\biggr],\\
&\alpha_3=\frac{\nu v_F^2 \alpha_R(\left[\mathbf{h} \times \hat{\mathbf{q}}\right] \cdot \mathbf{c})}{256\kappa^2\left(\pi T\right)^4}\left[93\kappa^2\zeta\left(5\right)-14\zeta\left(3\right)-\operatorname{Re}\left(\psi^{(2)}\left(\frac{1}{2}+\frac{i \kappa}{2}\right)\right)\right],\\
&\beta_1=\frac{\nu \alpha_R(\left[\mathbf{h} \times \hat{\mathbf{q}}\right] \cdot \mathbf{c})}{56\kappa^4\left(\pi T\right)^4} \left[93\kappa^4\zeta\left(5\right)-28\kappa^2\zeta\left(3\right)
-16\operatorname{Re}\left(\psi\left(\frac{1}{2}+\frac{i \kappa}{2}\right)-\psi\left(\frac{1}{2}\right)\right)+8\kappa\operatorname{Re}\left(i\psi^{(1)}\left(\frac{1}{2}+\frac{i \kappa}{2}\right)\right)\right],\\
&\beta_3= -\frac{\nu \alpha_R v_F^2(\left[\mathbf{h} \times \hat{\mathbf{q}}\right] \cdot \mathbf{c})}{12288\kappa^6\left(\pi T\right)^6} \left[17145\kappa^6\zeta\left(7\right)-1488\kappa^4\zeta\left(5\right)-336\kappa^2\zeta\left(3\right)+384\operatorname{Re}\left(\psi\left(\frac{1}{2}+\frac{i \kappa}{2}\right)-\psi\left(\frac{1}{2}\right)\right)\right. \nonumber \\ &\left.-192\kappa\operatorname{Re}\left(i\psi^{(1)}\left(\frac{1}{2}+\frac{i \kappa}{2}\right)\right)-72\kappa^2\operatorname{Re}\left(\psi^{(2)}\left(\frac{1}{2}+\frac{i \kappa}{2}\right)\right)+16\kappa^3\operatorname{Re}\left(i\psi^{(3)}\left(\frac{1}{2}+\frac{i \kappa}{2}\right)\right)\right],
\end{align}
\end{subequations}
\end{widetext}


\section{Calculations in the helical basis}\label{AppE}

\subsection{Helical Basis}

The helical basis is the basis that diagonalized the normal state Hamiltonian:
\begin{equation}
H=\sum_{\mathbf{k}ss'}\left[\varepsilon(\mathbf{k})\delta_{ss'}+\mathbf{g}(\mathbf{k})\cdot\boldsymbol{\sigma}_{ss'}\right]c^\dagger_{\mathbf{k}s}c_{\mathbf{k}s'}=\sum_{\mathbf{k}\lambda}\xi_\lambda(\mathbf{k})d^\dagger_{\mathbf{k}\lambda}d_{\mathbf{k}\lambda}
\end{equation}
where \(s,s'=\uparrow,\downarrow\) (also \(\pm1\) respectively when convenient below) are spin indices, $\varepsilon(\mathbf{k})=\frac{k^2}{2m}-\mu$ is the usual dispersion and
$\mathbf{g}(\mathbf{k})=\boldsymbol{\beta}(\mathbf{k})+\mathbf{h}$ 
includes the spin-orbit coupling (SOC) \(\boldsymbol{\beta}(\mathbf{k})=-\boldsymbol{\beta}(-\mathbf{k})\) and the Zeeman term with magnetic field \(\mathbf{h}\) measured in units of the Bohr magneton. For Rashba SOC in 2D we take $\boldsymbol{\beta}(\mathbf{k})=\alpha_R [\mathbf{p}\times \hat{\mathbf{z}}]$. The Hamiltonian is diagonalized by the operators (Einstein summation convention assumed below unless otherwise stated)
\[d_{\mathbf{k}\lambda}=U^s_\lambda(\mathbf{k})c_{\mathbf{k}s}\]
that form the so-called helical basis (also sometimes the SOC basis) with helical indices \(\lambda=\pm1\), and where
\[U^s_\lambda(\mathbf{k})=\frac{1}{\sqrt2}\sqrt{1+s\lambda\frac{g_z(\mathbf{k})}{g(\mathbf{k})}}\left(\lambda e^{-i\varphi(\mathbf{k})}\right)^{\frac{1+s}{2}}\]
where \(g(\mathbf{k})=|\mathbf{g}(\mathbf{k})|\) and
\[e^{i\varphi(\mathbf{k})}=\frac{g_x(\mathbf{k})+ig_y(\mathbf{k})}{\sqrt{g_x^2(\mathbf{k})+g_y^2(\mathbf{k})}}\]
For Rashba SOC, this simplifies to
\[U^s_\lambda(\mathbf{k})=\frac{1}{\sqrt2}\left(i \lambda e^{-i\theta_\mathbf{k}}\right)^{\frac{1+s}{2}}\]
where \(\theta_\mathbf{k}\) is just the angle of \(\mathbf{k}\). In general,
\[\xi_\lambda(\mathbf{k})=\varepsilon(\mathbf{k})+\lambda g(\mathbf{k})\]

\subsection{Gap function and interactions in the helical basis}

The gap function can also be re-written in the helical basis. To do this, we note that

\begin{align}
\sum_{\mathbf{k}ss'}\widehat{\Delta}_{ss'}(\mathbf{k;q})c^\dagger_{\mathbf{k+q}/2,s} c^\dagger_{-\mathbf{k+q}/2,s'}=\nonumber \\
\sum_{\mathbf{k}\lambda\lambda'}\widehat{\Delta}_{\lambda\lambda'}(\mathbf{k;q})d^\dagger_{\mathbf{k+q}/2,\lambda} d^\dagger_{-\mathbf{k+q}/2,\lambda'}
\end{align}
so
\[\widehat{\Delta}_{\lambda\lambda'}(\mathbf{k;q})=U^s_\lambda(\mathbf{k+q}/2)U^{s'}_{\lambda'}(-\mathbf{k+q}/2)\widehat{\Delta}_{ss'}(\mathbf{k;q})\]
with sum over \(s,s'\) on the RHS implied. Writing out
\[\widehat{\Delta}_{ss'}(\mathbf{k;q})=d_\mu(\mathbf{k;q})(\sigma^\mu i\sigma^y)_{ss'}\]
it is convenient to define
\[\widehat{\Delta}_{\lambda\lambda'}(\mathbf{k;q})=d_\mu(\mathbf{k;q})W^{(\mu)}_{\lambda\lambda'}(\mathbf{k;q})\]
with
\[W^{(\mu)}_{\lambda\lambda'}(\mathbf{k;q})=\sum_{ss'}(\sigma^\mu i\sigma^y)_{ss'}U^s_\lambda(\mathbf{k+q}/2)U^{s'}_{\lambda'}(-\mathbf{k+q}/2)\]
To leading order it is usually sufficient to take \(\mathbf{q}=0\), since higher order terms in \(q\) go as \(g/v_F\), which is generally very small. In the absence of magnetic field and general SOC, we then have
\begin{widetext}
\begin{align}
    W^{(0)}(\mathbf{k;0})&=e^{-i\varphi(\mathbf{k})}\varsigma^z\nonumber\\
    W^{(x)}(\mathbf{k;0})&=e^{-i\varphi(\mathbf{k})}\left[\frac{g_x(\mathbf{k})}{g(\mathbf{k})}\varsigma^0-\frac{g_x(\mathbf{k})g_z(\mathbf{k})}{g^2(\mathbf{k})}i\varsigma^y+\frac{g_y(\mathbf{k})}{g(\mathbf{k})}i\varsigma^x\right]\nonumber\\
    W^{(y)}(\mathbf{k;0})&=e^{-i\varphi(\mathbf{k})}\left[\frac{g_y(\mathbf{k})}{g(\mathbf{k})}\varsigma^0-\frac{g_y(\mathbf{k})g_z(\mathbf{k})}{g^2(\mathbf{k})}i\varsigma^y+\frac{g_x(\mathbf{k})}{g(\mathbf{k})}i\varsigma^x\right]\nonumber\\
    W^{(z)}(\mathbf{k;0})&=e^{-i\varphi(\mathbf{k})}\left[\frac{g_z(\mathbf{k})}{g(\mathbf{k})}\varsigma^0+\sqrt{1-\frac{g_z^2(\mathbf{k})}{g^2(\mathbf{k})}}i\varsigma^y\right]
\end{align}
where \(\varsigma^\mu\) are Pauli matrices with helical indices. The pairing assumed in \cite{IlicBergeret22} is exactly \(W^{(0)}(\mathbf{k;0})\), which is the same as the pure s-wave singlet pairing in the absence of magnetic field and finite momentum pairing assumed in \cite{YuanFu22}. Note that in the strong SOC limit the off-diagonal terms can be dropped.

Working to linear order in \(h\), we can also obtain expressions for \(\mathbf{h}\) applied along the \(\hat{\mathbf{x}}\) direction, assuming \(\mathbf{g}\) is not out of plane. It turns out that corrections to \(W^{(x)}\) and \(W^{(y)}\) are of order \(h/\sqrt{g^2-g_z^2}\), and
\begin{align}
    W^{(0)}(\mathbf{k;0})&=e^{-i\varphi(\mathbf{k})}\left[\varsigma^z+\frac{h\sin\theta_\mathbf{k}}{\sqrt{g^2(\mathbf{k})-g_z^2(\mathbf{k})}}\left(i\varsigma^y+\frac{g_z(\mathbf{k})}{g(\mathbf{k})}\varsigma^{0}\right)\right]\nonumber\\
    W^{(z)}(\mathbf{k;0})&=e^{-i\varphi(\mathbf{k})}\left[\frac{g_z(\mathbf{k})}{g(\mathbf{k})}\varsigma^0+\sqrt{1-\frac{g_z^2(\mathbf{k})}{g^2(\mathbf{k})}}i\varsigma^y+\frac{h\sin\theta_\mathbf{k}}{\sqrt{g^2(\mathbf{k})-g_z^2(\mathbf{k})}}\varsigma^{z}\right]
\end{align}
Phase \(\varphi\) here is to be evaluated at \(h=0\); for \(\mathbf{h}\) along \(y\), simply replace \(\sin\theta_\mathbf{k}\) with \(-\cos\theta_\mathbf{k}\).

The same transformation can be applied to pairing interactions:
\begin{align}
    H_{int}&=\sum V^{(\mu)}(\mathbf{p,k;q}) (\sigma^\mu i\sigma^y)^\dagger_{s_1s_2}(\sigma^\mu i\sigma^y)_{s_3s_4}c^\dagger_{\mathbf{p}s_3}c^\dagger_{-\mathbf{p}s_4}c_{\mathbf{k}s_1}c_{-\mathbf{k}s_2}=\nonumber\\
    &=\sum V^{(\mu)}(\mathbf{p,k;q}) W^{(\mu)\dagger}_{\lambda_1\lambda_2}(\mathbf{k})W^{(\mu)}_{\lambda_3\lambda_4}(\mathbf{p})d^\dagger_{\mathbf{p}\lambda_3}d^\dagger_{-\mathbf{p}\lambda_4}d_{\mathbf{k}\lambda_1}d_{-\mathbf{k}\lambda_2}
\end{align}
\end{widetext}
(with analogous expressions for Dzyaloshinskii–Moriya interactions that involve \((\sigma^\mu i\sigma^y)^\dagger_{s_1s_2}(\sigma^\nu i\sigma^y)_{s_3s_4}\) terms with \(\mu\neq\nu\)). One particular consequence of this is that in the absence of the magnetic field the \(s\)-wave singlet channel remains the self-consistent solution of the linearized gap equation and decouples from the triplet channels if the interactions are of on-site Hubbard type: the relevant traces \(\text{Tr}\left[W^{(0)\dagger}W^{(j)}\right]=0\) vanish for \(j=x,y,z\) (this implies an emergent/accidental symmetry). This is no longer the case already in the slightly extended Hubbard model with uniform interactions on each helical band, but with unequal interband and intraband interactions.

\subsection{Corrections to form factors due to magnetic field and finite-momentum pairing}

Here we consider corrections to the form factors for the \(s\)-wave spin singlet case due to magnetic field and finite-momentum pairing, including also the interband pairing. We compute \(W^{(0)}(\mathbf{k;q})\) to all orders in \(\mathbf{h}\) and \(\mathbf{q}\) for the special case of \(g_z=0\). The result is
\begin{widetext}
\[W^{(0)}(\mathbf{k;q})=\frac{1}{2}\left[\varsigma^z\left(e^{-i\varphi(\mathbf{k;q}/2)}-e^{-i\varphi(-\mathbf{k;q}/2)}\right)+i\varsigma^y\left(e^{-i\varphi(\mathbf{k;q}/2)}+e^{-i\varphi(-\mathbf{k;q}/2)}\right)\right]\]
where
\[e^{-i\varphi(\mathbf{k;q}/2)}=\frac{b_x+\alpha_R p_y-i(b_y-\alpha_R p_x)}{\sqrt{(b_x+\alpha_R p_y)^2+(b_y-\alpha_R p_x)^2}}\]
where for convenience we introduce the vector \(\mathbf{b}=\mathbf{h}-\alpha_R[\mathbf{q}\times\hat{\mathbf{z}}]/2\). The expressions for the GL coefficients now become
\begin{align}
\alpha(\mathbf{q})&=\left[\frac{1}{2}\sum_{\mathbf{k}\lambda\lambda'}\left|W^{(0)}_{\lambda\lambda'}(\mathbf{k;q})\right|^2\Pi_{\lambda\lambda'}(\mathbf{k;q})\right]^{T_c,b=0}_T\\
\beta(\mathbf{q})&=-\frac{1}{4}\sum_{\mathbf{k}\lambda_1\dots\lambda_4}W^{(0)*}_{\lambda_1\lambda_2}(\mathbf{k;q})W^{(0)}_{\lambda_3\lambda_2}(\mathbf{k;q})W^{(0)*}_{\lambda_3\lambda_4}(\mathbf{k;q})W^{(0)}_{\lambda_4\lambda_1}(\mathbf{k;q})\Gamma_{\lambda_1\lambda_2\lambda_3\lambda_4}(\mathbf{k;q})
\end{align}
where, working to first order in \(g/v_F\) and assuming a dispersion linearized about the Fermi surface,
\(\xi_\lambda(\mathbf{k+q}/2)\approx v_{\lambda}(\theta_\mathbf{k})\delta k+\mu_\lambda(\theta_\mathbf{k};\mathbf{q})\), we have
\begin{align}
\bar{\Pi}_{\lambda\lambda'}(\theta;\mathbf{q})&\approx \nu_{\lambda\lambda'}\left[-\ln\frac{1.13\Lambda}{T}+\digamma\left(\hat{\mu}_{\lambda'\lambda},T\right)\right]\label{Pi}\\
\bar{\Gamma}_{\lambda_1\lambda_2\lambda_3\lambda_4}(\theta_\mathbf{k};\mathbf{q})&\approx \sum_{j=1}^4 \frac{\nu_{\lambda_j\lambda_{j+1}}\digamma\left(\hat{\mu}_{\lambda_{j+1}\lambda_j},T\right)}{(v_j\mu_{j+1,j+2}-v_{j+1}\bar{\mu}_{j+2,j}-v_{j+2}\mu_{j,j+1})(v_j\bar{\mu}_{j+1,j+3}-v_{j+1}\mu_{j+3,j}+v_{j+3}\mu_{j+1,j})}\label{Gamma}
\end{align}
\end{widetext}
where \(j\) is defined modulo four, \(\Lambda\) is some high energy cutoff,
\begin{align}
    \nu_{\lambda\lambda'}&=\frac{2\nu_{\lambda} \nu_{\lambda'}}{\nu_{\lambda}+\nu_{\lambda'}}\\
    \hat{\mu}_{\lambda'\lambda}(\theta_\mathbf{k};\mathbf{q})&=\frac{\nu_{\lambda'}\mu_{\lambda'}(\theta_\mathbf{k};\mathbf{q})-\nu_{\lambda}\mu_{\lambda}(\theta_\mathbf{-k};\mathbf{q})}{\nu_{\lambda'}+\nu_{\lambda}}
\end{align}
are the `reduced' DOSs and nesting detuning parameters respectively, and 
\[\digamma\left(x,T\right)=\text{Re}\left[\psi\left(\frac{1}{2}+\frac{i x}{2\pi T}\right)-\psi\left(\frac{1}{2}\right)\right]\label{varpiPDW}\]
where \(\psi(x)\) is the digamma function. In Eq. \ref{Gamma} one can replace \(\digamma\left(x,T\right)\) with simply \(\psi\left(\frac{1}{2}+\frac{i x}{4\pi T}\right)\). Here we also defined the shorthand \(\mu_{jj'}=\mu_{\lambda_j}+\mu_{\lambda_j'}\) and \(\bar{\mu}_{jj'}=\mu_{\lambda_j}-\mu_{\lambda_j'}\).

\subsection{Logarithmic Corrections to the Lifshitz Invariant}

As noted in \cite{AttiasMichaeliKhodas23}, there are logarithmic corrections to the Lifshitz invariant \(\alpha_1\) in the presence of an external magnetic field and non-zero Cooper pair momentum and that need to be treated more carefully. Computing \(\alpha(\mathbf{q})\), we have
\begin{widetext}
\[\frac{1}{2}\sum_{\lambda\lambda'}\left|W^{(0)}_{\lambda\lambda'}(\mathbf{k;q})\right|^2\Pi_{\lambda\lambda'}(\mathbf{k;q})=\frac{1-\cos(\varphi_\mathbf{p}-\varphi_{-\mathbf{p}})}{4}(\Pi_{++}+\Pi_{--})+\frac{1+\cos(\varphi_\mathbf{p}-\varphi_{-\mathbf{p}})}{4}(\Pi_{+-}+\Pi_{-+})\]    
with the shorthand 
\(\varphi_\mathbf{p}=\varphi(\mathbf{p;q})\). Note that
\[\cos(\varphi_\mathbf{p}-\varphi_{-\mathbf{p}})=\frac{b^2-(\alpha_Rp)^2}{\sqrt{(b^2+(\alpha_Rp)^2)^2-4\alpha^2_R|\mathbf{b}\times\mathbf{p}|^2}}\]
so at \(b=0\), \(\cos(\varphi_\mathbf{p}-\varphi_{-\mathbf{p}})=-1\).

Let us now assume \(h\ll \alpha_R p_F\). We then have
\[W^{(0)}(\mathbf{p;q})\approx ie^{-i\theta_\mathbf{p}}\left(\varsigma^z\left(1-\frac{\mathbf{b}\cdot\hat{\mathbf{p}}\,(\mathbf{b}\cdot\hat{\mathbf{p}}-2 i(\mathbf{b}\times\hat{\mathbf{p}})_z)}{2(\alpha_R p_F)^2}\right)+\varsigma^y \frac{\mathbf{b}\cdot\hat{\mathbf{p}}}{\alpha_R p_F}\right)\]
such that
\begin{align}
    \frac{1}{2}\sum_{\lambda\lambda'}\left|W^{(0)}_{\lambda\lambda'}(\mathbf{k;q})\right|^2\Pi_{\lambda\lambda'}(\mathbf{k;q})&=\frac{1}{2}\left(1-\left(\frac{\mathbf{b}\cdot\hat{\mathbf{p}}}{\alpha_R p_F}\right)^2\right)(\Pi_{++}+\Pi_{--})+\frac{1}{2}\left(\frac{\mathbf{b}\cdot\hat{\mathbf{p}}}{\alpha_R p_F}\right)^2(\Pi_{+-}+\Pi_{-+})
\end{align}
(note that \(|W^{(0)}|^2=1\)). The relevant detuning parameters are, to second and first order in \(b\) respectively,
\begin{align}
    \hat{\mu}_{\lambda\lambda}(\mathbf{p;q})&=\frac{\xi_\lambda(\mathbf{p+q}/2)-\xi_\lambda(-\mathbf{p+q}/2)}{2}=\frac{\mathbf{p\cdot q}}{2m}+\lambda(\mathbf{b}\times\hat{\mathbf{p}})_z=\frac{\mathbf{v}_{\lambda}\cdot\mathbf{q}}{2}+\lambda(\mathbf{h}\times\hat{\mathbf{p}})_z\\
    \hat{\mu}_{\lambda,-\lambda}(\mathbf{p;q})&=\frac{\nu_\lambda\xi_\lambda(\mathbf{p+q}/2)-\nu_{-\lambda}\xi_{-\lambda}(-\mathbf{p+q}/2)}{2\nu}=\frac{\mathbf{p\cdot q}}{2m}+\lambda \alpha_R p+\frac{\alpha_R}{v_F}(\mathbf{b}\times\hat{\mathbf{p}})_z
\end{align}
\end{widetext}
where \(\mathbf{v}_{\lambda}(\mathbf{p})=\mathbf{p}/m+\lambda\alpha_R\mathbf{\hat{p}}\).

Let us now further consider the case \(h\ll\alpha_R p_F\ll T_c\). In that case
\[\Pi_{\pm,\mp}\approx \nu_{\lambda,-\lambda}\left[-\ln\frac{1.13\Lambda}{T}+\frac{7 (\alpha_R p_F)^2 \zeta(3)}{4 \pi^2 T^2}\right]\]
while
\[(\Pi_{++}+\Pi_{--})/2\approx \alpha^{(0)}(\mathbf{q})\]
where \(\alpha^{(0)}\) is \(\alpha(\mathbf{q})\) found by taking \(\mathbf{h}=\mathbf{q}=0\) in \(W^{(0)}\). The logarithmic correction to \(\alpha(\mathbf{q})\) is then
\begin{align}
    \delta\alpha_{\log}&\approx -\frac{\Delta \nu^2}{\nu} \ln\frac{\Lambda}{T_c}\int \left(\frac{\mathbf{b}\cdot\hat{\mathbf{p}}}{\alpha_R p_F}\right)^2\frac{d\theta_\mathbf{p}}{2\pi}\\
    &\approx \frac{\Delta \nu^2}{\nu} \ln\frac{\Lambda}{T_c}\frac{h q/p_F \sin\theta_\mathbf{q}+\alpha_R p_F(q/p_F)^2/4}{2\alpha_R p_F}\nonumber
\end{align}
where in the last expression we dropped the \(h^2\) term. The leading correction to \(\alpha(\mathbf{q})\) is therefore of order \(\sim \nu\delta^2(h/\alpha_R p_F)\sin\theta_\mathbf{q}\ln(\Lambda/T_c)\), which is beyond the linear order in \(\delta\) we are considering. In this limit the logarithmic correction can therefore be neglected.

Of more interest is the case \(h\ll T_c\ll\alpha_R p_F\), for which we have
\[\Pi_{\pm,\mp}\approx -\nu_{\lambda,-\lambda}\ln\frac{\Lambda}{\alpha_R p_F}\]
i.e. the logarithm is cut by \(\alpha_R p_F\) instead of \(T_c\), as expected. The logarithmic correction to \(\alpha(\mathbf{q})\) is then
\begin{align}
    \delta\alpha_{\log}&\approx \frac{\nu}{2}\left( \ln\frac{\Lambda}{T_c}- \ln\frac{\Lambda}{\alpha_R p_F}\right)\int \left(\frac{\mathbf{b}\cdot\hat{\mathbf{p}}}{\alpha_R p_F}\right)^2\frac{d\theta_\mathbf{p}}{2\pi}\approx\nonumber\\
    &\approx \frac{\nu}{2} \ln\frac{\alpha_R p_F}{T_c}\frac{h q/p_F \sin\theta_\mathbf{q}+\alpha_R p_F(q/p_F)^2/4}{2\alpha_R p_F}
\end{align}
where we dropped terms of order \(h^2\) and \(\delta^2\) in the last line. This means, in particular, that the computation of \(\alpha_1\) presented in Ref. \cite{IlicBergeret22} is strictly speaking valid only for \(h (q/p_F)\ll \alpha_R p_F /\ln\frac{\alpha_R p_F}{T_c} \ll \alpha_R p_F\), which can be much more stringent than \(h \ll \alpha_R p_F\), depending on the value of \((q/p_F)\ln\frac{\alpha_R p_F}{T_c}\).

The effect on the superconducting diode coefficient \(\eta\), however, is more benign. We note, first, that for Rashba SOC, to leading order \(\alpha_2\beta_2=4\alpha_4\beta_0\) (that is at \(h=0\)), and as a result the terms proportional to \(\alpha_1\) in Eq. (\ref{etaGL}) cancel out. As a consequence, changing only the Lifshitz invariant does not modify \(\eta\), another special feature of the \(s\)-wave Rashba superconductor in the strong SOC limit. Consequently, the logarithmic correction to \(\alpha_2\) must also be included to get a non-zero correction to \(\eta\), which is
\[\delta\eta=-\frac{186\, \sqrt{2}\, \zeta(5)}{
 343\, \sqrt{7\zeta^{7}(3)}}\left(\frac{\delta}{\kappa}\right)^3\ln^2\frac{\alpha_R p_F}{T_c} \frac{h\sqrt{t}}{\alpha_Rp_F} \sin\theta_\mathbf{q}\]
This is to be compared to Eq. \eqref{etaLargeSOC}, which gives \(\eta\propto \delta h\sqrt{t}/T/\kappa^2\). In particular, since we are working to linear order in \(\delta\), the logarithmic correction to \(\eta\) is therefore negligible and heavily suppressed by the Fermi energy. Computing the \(h^3\) correction, on the other hand, we find (to leading order in \(h/(\alpha_Rp_F)\))
\begin{align}
 \delta\eta&=\frac{635\,  \zeta(7)}{
 56 \sqrt{42\zeta^{3}(3)} }\frac{\delta\, h^3 \sqrt{t}\, \sin^3\theta_\mathbf{q}}{\pi^3T_c^3}\approx\nonumber\\
 &\approx0.0432 \frac{\delta\, h^3 \sqrt{t}\, \sin^3\theta_\mathbf{q}}{T_c^3}.
\end{align}


%

\end{document}